\documentclass[a4paper,11pt]{article}

\usepackage[utf8]{inputenc}
\usepackage{amsmath}
\usepackage{tikz}
\usepackage{verbatim}
\usepackage{pgffor}
\usepackage{listings}
\usepackage{placeins}
\usepackage{float}
\usepackage{cite}
\usepackage{algorithm,algorithmic}
\usepackage{color}
\usepackage{jheppub}
\usepackage{tcolorbox}
\usepackage[thinlines]{easytable}
\usepackage{cite}
\usepackage{hyperref}
\usetikzlibrary{decorations.pathmorphing}
\usetikzlibrary{shapes.geometric}

\tikzset{snake it/.style={decorate, decoration=snake}}
\tikzset{wavy/.style={decorate, decoration={complete sines,amplitude=1.5pt, segment length=4pt}}}
\usetikzlibrary{matrix}
\usetikzlibrary{arrows.meta}
\usetikzlibrary{shapes.geometric, arrows}
\tikzstyle{startstop} = [rectangle, rounded corners, minimum width=3cm, minimum height=1cm,text centered, draw=black, fill=red!30]
\tikzstyle{io} = [trapezium, trapezium left angle=70, trapezium right angle=110, minimum width=3cm, minimum height=1cm, text centered, draw=black, fill=blue!30]
\tikzstyle{process} = [rectangle, minimum width=3cm, minimum height=1cm, text centered, draw=black, fill=orange!30]
\tikzstyle{decision} = [diamond, minimum width=3cm, minimum height=1cm, text centered, draw=black, fill=green!30]
\tikzstyle{arrow} = [thick,->,>=stealth]

\newcommand{\dlog}{{\it d}{\textup log}}
\newcommand{\dlogt}{{\it d}{\textup log}\;}

\newcommand{\beq}{\begin{equation}}
\newcommand{\eeq}{\end{equation}}

\makeatletter
\def\maketag@@@#1{\hbox{\m@th\normalfont\normalsize#1}}
\makeatother

\def\eps{\epsilon}

\title{Constructing d-log integrands and computing master integrals for three-loop four-particle scattering}
\author[a]{Johannes Henn}
\author[b]{Bernhard Mistlberger}
\author[c]{Vladimir A. Smirnov}
\author[d]{Pascal Wasser}

\affiliation[a]{Max-Planck-Institut f\"{u}r Physik, Werner-Heisenberg-Institut, D-80805 M\"{u}nchen, Germany}
\affiliation[b]{Center for Theoretical Physics, Massachusetts Institute of Technology, Cambridge, MA 02139, USA}
\affiliation[c]{Skobeltsyn Institute of Nuclear Physics of Moscow State University 119991, Moscow, Russia}
\affiliation[d]{PRISMA+ Cluster of Excellence, Johannes Gutenberg University, D-55099 Mainz, Germany}

\emailAdd{henn@mpp.mpg.de}
\emailAdd{bernhard.mistlberger@gmail.com}
\emailAdd{smirnov@theory.sinp.msu.ru}
\emailAdd{wasserp@uni-mainz.de}

\preprint{MPP-2020-13, MITP/20-006, MIT-CTP/5176}

\bigskip
\abstract{
We compute all master integrals for massless three-loop four-particle scattering amplitudes required for processes like di-jet or di-photon production at the LHC.
We present our result in terms of a Laurent expansion of the integrals in the dimensional regulator up to 8$^{\text{th}}$ power, with coefficients expressed in terms of harmonic polylogarithms. 
As a basis of master integrals we choose integrals with integrands that only have logarithmic poles - called \dlogt forms.
This choice greatly facilitates the subsequent computation via the method of differential equations.
We detail how this basis is obtained via an improved algorithm originally developed by one of the authors.
We provide a public implementation of this algorithm. 
We explain how the algorithm is naturally applied in the context of unitarity.
In addition, we classify our \dlogt forms according to their soft and collinear properties.
}

\begin{document}

\setcounter{tocdepth}{2}
\maketitle
\setcounter{page}{1}

\clearpage
\section{Introduction}

Perturbative quantum field theory allows us to derive predictions for physical observable from our in-principle understanding of the fundamental interactions of nature. 
Experiments like the Large Hadron Collider (LHC) allow us to measure such observables and test our current conceptions of the world. 
One particular observable that allows us to probe the strong interactions is the production cross section of sprays of collimated hadrons -- so-called jets. 
This observable is measured with astounding precision at the LHC. 
Consequently, in order to maximally benefit from this measurement the precision of the theoretical prediction for this observable must at least match the experimental one.
In order to achieve this goal it is necessary to compute sufficiently many orders in the perturbative expansion of the cross section for the desired observable.

When physical quantities in quantum field theory are expanded perturbatively
in the coupling constant, corrections beyond the leading order involve Feynman loop integrals.
Examples are correlation functions depending on positions of operators, or scattering amplitudes 
depending on on-shell particle momenta. Feynman integrals typically evaluate to multi-valued
functions, such as logarithms, dilogarithms, and generalizations thereof.

It is of great physical but also mathematical interest to understand better the connection between
the Feynman integrals and the special functions that arise.
In recent years, such insights allow us to predict the type of special functions, and their `fine structure',
that arise from carrying out the loop integrations, simply by analyzing properties of the Feynman {\it integrand}.
These insights have already had numerous applications and streamlined many complicated calculations.

An important class of special functions is that of multiple polylogarithms~\cite{Goncharov:1998kja,Chen:1977oja}. 
They are iterated integrals having the same integration kernels as logarithms.  
The number of integrations of multiple polylogarithms is called the (transcendental) weight.
For example, logarithm and dilogarithm have weight one and two, respectively.
Functions with more general integration kernels may also arise in Feynman integrals, but are beyond the scope of the present paper and are not discussed here.

A heuristic observation is that $L$-loop integrals in four dimensions give rise to functions
of weight lower or equal to $2 \, L $. For example, at one loop in four dimensions,  the maximal weight is two,
which means that the space of functions is given by algebraic functions, (products of) logarithms, and dilogarithms.
A special role is played by the functions of maximal weight $2 L $.
Many examples of such functions were encountered in $\mathcal{N}=4$ supersymmetric Yang-Mills theory.
It appears that many quantities in that theory are naturally expressed in terms of functions of uniform and 
maximal transcendental weight, see e.g. \cite{Kotikov:2004er,Henn:2016jdu,Abreu:2018aqd,Caron-Huot:2019vjl,Badger:2019djh}.

There is a conjectured connection between uniform weight integrals and properties of their integrands:
the singularities of the integrand are locally of logarithmic type.
This conjecture has been tested for many cases, originally in the context of planar, 
finite integrals in $\mathcal{N}=4$ supersymmetric Yang-Mills theory. However, this notion generalizes in a number
of ways. First of all, the dual conformal symmetry of the theory (which implies a certain power counting)
is not essential: for example, at one loop both box and triangle integrals give rise to uniform weight functions.
Moreover, generalizations include non-planar integrals, integrals involving massive particles, for example.
An important further generalization concerns dimensional regularization, where integrals are computed in $D=4-2 \eps$ 
dimensions, in a Laurent expansion for small $\eps$. Observing that poles such as $1/\eps$ in the dimensional
regulator would correspond to $\log \Lambda$ for some cutoff $\Lambda$, it is natural to assign a transcendental weight $-1$ to $\eps$.
This seemingly simple concept has important repercussions. What does it mean for a function $f(\eps,x)$ to have uniform weight?
Writing 
\begin{align}\label{purefunction1}
f(\eps,x) = \frac{1}{\eps^{2 L}} \sum_{k\ge 0} \eps^k f^{(k)}(x)\,,
\end{align}
it means that $f^{(k)}(x)$ has weight $k$, for any order in the expansion!
This is a rather strong condition.

In practice, the fact that properties of the loop integrand may predict which integrals evaluate to uniform weight functions
is extremely helpful.  The classification of integrands having \dlogt forms can be done at the integrand level, i.e. prior
to integration. 
This connection is well-known and has been investigated and used in a number of papers, e.g. 
\cite{ArkaniHamed:2010gh,Gehrmann:2011xn,Drummond:2013nda,Arkani-Hamed:2014via,Bern:2014kca}.
An algorithm to do this was implemented in \cite{WasserMSc}. 
It is based on a suitable parametrization of the loop integrand, and analyzes the resulting rational function by taking residues iteratively.
This approach is complementary to the algorithm implemented in \cite{Larsen:2017kzf} that uses methods from computational algebraic geometry to compute multivariate residues.
Algorithms that can be applied to Feynman integrals (in contrast to integrands) in conjunction with the methods of differential equations~\cite{Kotikov:1990kg,Kotikov:1991hm,Kotikov:1991pm,Henn:2013pwa,Gehrmann:1999as} in order to find uniform weight integrals were discussed in refs.~\cite{Gituliar:2017vzm,Lee:2014ioa,Meyer:2016slj,Prausa:2017ltv,Dlapa:2020cwj,Hoschele:2014qsa}

In the present paper, we discuss a refined version of the algorithm of ref.~\cite{WasserMSc} to find \dlogt forms. 
 The improvements mainly concern the following two points. 
 Firstly, at some stage of taking residues, one may encounter integrands with denominators that are quadratic in the integration variables. 
 We introduce a method that allows the algorithm to proceed in those cases.
 Secondly,  the analysis performed to find integrands having \dlogt forms is closely related to taking
(generalized) cuts of integrands, and in particular to leading singularities.
The latter correspond to taking the loop integrand, and performing contour integrals to take multiple residues,
thereby completely localizing the integration. Obviously, doing so is much simpler than carrying out the
loop integration over Minkowski space-time. 
  We use this connection to organize the analysis of loop integrands according to
 different cuts, thereby simplifying each individual calculation.

It is worth pointing out that generalized cuts and leading singularities are also important methods
for computing loop integrands that bypass Feynman diagrams. 
Given the way it is defined, the uniform weight integrands we construct are very natural building blocks
for such integrand constructions, and we expect our results to be useful in this area. 
For recent references in this direction, see e.g. \cite{Kosower:2011ty,Badger:2012dv,Mastrolia:2012an,Ita:2015tya,Bourjaily:2015jna,Bourjaily:2017wjl}. 

There is a further application of \dlogt integrands, namely an improved control over the singularities of Feynman integrals after integration.
On the one hand, it turns out that \dlogt integrals in four dimensions are ultraviolet finite. This can be shown by a power counting argument which we explain below. 
On the other hand, on-shell amplitudes may have  infrared (soft and collinear) divergences. For a given loop integrand, it is easy to analyze the soft and collinear
behaviour responsible for divergences. By doing so one may select a basis of loop \mbox{integrands}/integrals with improved convergence properties.
While examples of this are well known at one loop, this was first discussed systematically at higher loops in \cite{ArkaniHamed:2010gh},
with the aim of introducing finite loop integrands that are relevant for infrared-finite parts of scattering processes.
The improved understanding of infrared properties of loop integrands was also used to determine the latter via bootstrap methods \cite{Drummond:2010mb,Bourjaily:2011hi}. 
See ref.~\cite{Henn:2019rmi} for a recent application of the classification of \dlogt intergrands according to divergence structure to four-loop form factors.
It is possible to algorithmically find finite but not necessarily uniform transcendental Feynman integrals, see for example refs.~\cite{vonManteuffel:2014qoa,vonManteuffel:2015gxa}.

Let us now return to the question of the evaluation of the loop integrals.
As was already mentioned, knowing (conjecturally) that a given loop integrand integrates to a pure uniform weight function
provides a lot of information. In fact, it is easy to see that a pure function satisfies simple differential equations.
Moreover, any Feynman integral satisfies some $n$-th order differential equation. 
Equivalently, one may transform this into an $n\times n$ system of first-order differential
equations for the Feynman integral and other functions (e.g. derivatives).
Combining this with the information about the form of differential equations for pure functions
one may conclude that one may always reach a canonical form of the differential equations \cite{Henn:2013pwa}.
The latter are very useful for computing Feynman integrals, as they are in a form where the solution
in terms of special functions can directly be read off.

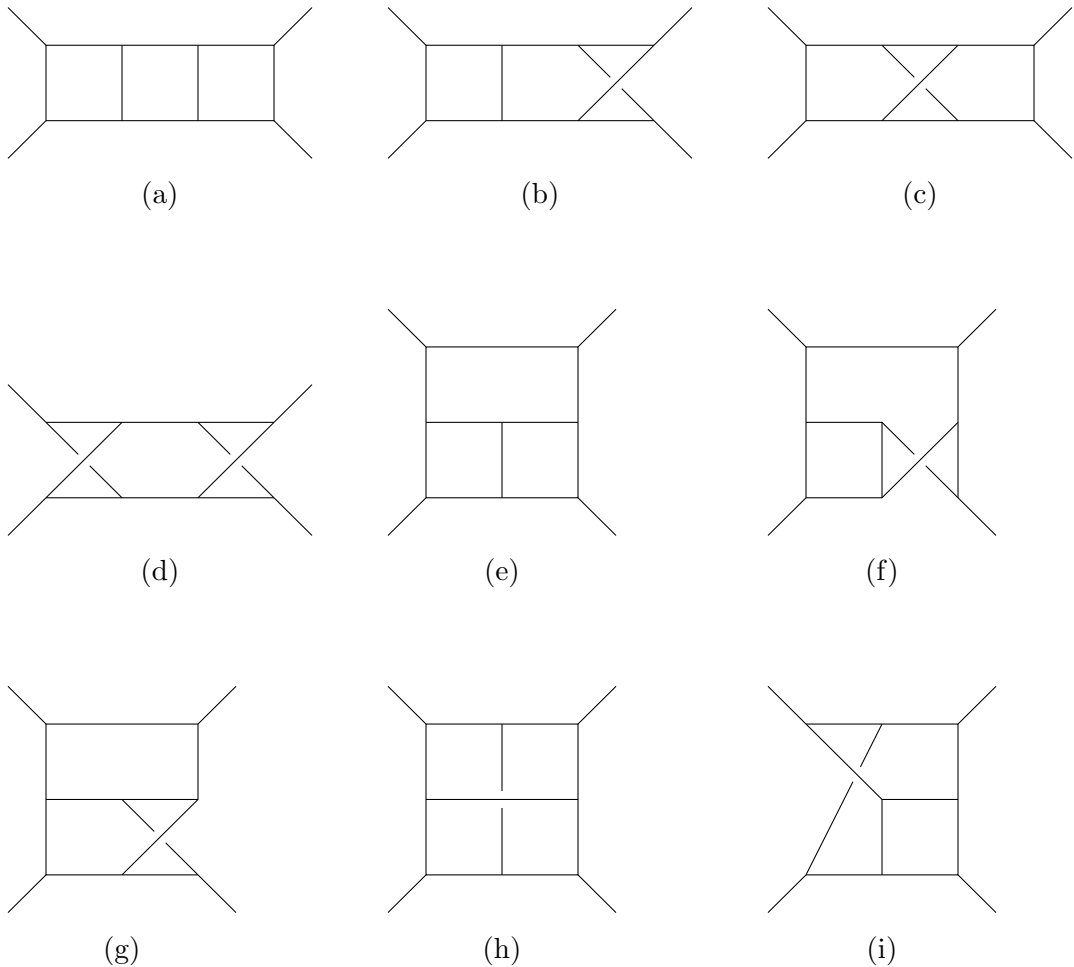
\begin{figure}[t]
\centering
\begin{tikzpicture}[scale=.5]
\def\gap{0.15};
\begin{scope}[scale=1.]
  \draw (0,0) -- (6,0) -- (6,2) -- (0,2) -- (0,0);
  \draw (2,0) -- (2,2);
  \draw (4,0) -- (4,2);
  \draw (0,0) -- (-1,-1);
  \draw (6,0) -- (7,-1);
  \draw (6,2) -- (7,3);
  \draw (0,2) -- (-1,3); 
  \draw (3,-2) node {(a)};
\end{scope}
\begin{scope}[shift={(10,0)}]
  \draw (0,0) -- (6,0);
  \draw (6,2) -- (0,2) -- (0,0);
  \draw (2,0) -- (2,2);
  \draw (4,0) -- (6,2);
  \draw (6,0) -- (5+\gap,1-\gap);
  \draw (5-\gap,1+\gap) -- (4,2);
  \draw (0,0) -- (-1,-1);
  \draw (6,0) -- (7,-1);
  \draw (6,2) -- (7,3);
  \draw (0,2) -- (-1,3); 
  \draw (3,-2) node {(b)};
\end{scope}
\begin{scope}[shift={(20,0)}]
  \draw (0,0) -- (6,0) -- (6,2) -- (0,2) -- (0,0);
  \draw (2,0) --  (4,2);
  \draw (4,0) -- (3+\gap,1-\gap);
  \draw (3-\gap,1+\gap) -- (2,2);
  \draw (0,0) -- (-1,-1);
  \draw (6,0) -- (7,-1);
  \draw (6,2) -- (7,3);
  \draw (0,2) -- (-1,3); 
  \draw (3,-2) node {(c)};
\end{scope}
\begin{scope}[shift={(0,-10)}]
  \draw (0,0) -- (6,0);
  \draw (6,2) -- (0,2);
  \draw (0,0) -- (2,2);
  \draw (2,0) -- (1+\gap,1-\gap);
  \draw (1-\gap,1+\gap) -- (0,2);
  \draw (4,0) -- (6,2);
  \draw (6,0) -- (5+\gap,1-\gap);
  \draw (5-\gap,1+\gap) -- (4,2);
  \draw (0,0) -- (-1,-1);
  \draw (6,0) -- (7,-1);
  \draw (6,2) -- (7,3);
  \draw (0,2) -- (-1,3); 
  \draw (3,-2) node {(d)};
\end{scope}
\begin{scope}[shift={(10,-10)}]
  \draw (0,0) -- (4,0) -- (4,4) -- (0,4) -- (0,0);
  \draw (0,2) --  (4,2);
  \draw (2,0) -- (2,2);
  \draw (0,0) -- (-1,-1);
  \draw (4,0) -- (5,-1);
  \draw (4,4) -- (5,5);
  \draw (0,4) -- (-1,5); 
  \draw (2,-2) node {(e)};
\end{scope}
\begin{scope}[shift={(20,-10)}]
  \draw (4,0) -- (4,4) -- (0,4) -- (0,0) ;
  \draw (4,0) -- (3+\gap,1-\gap);
  \draw (3-\gap,1+\gap) -- (2,2) -- (0,2);
  \draw (2,2) -- (2,0) ;
  \draw (0,0) -- (2,0) -- (4,2);
  \draw (0,0) -- (-1,-1);
  \draw (4,0) -- (5,-1);
  \draw (4,4) -- (5,5);
  \draw (0,4) -- (-1,5);
  \draw (2,-2) node {(f)}; 
\end{scope}
\begin{scope}[shift={(0,-20)}]
  \draw (4,4) -- (0,4) -- (0,0) -- (4,0) ;
  \draw (4,0) -- (3+\gap,1-\gap);
  \draw (3-\gap,1+\gap) -- (2,2);
  \draw (0,2) --  (4,2);
  \draw (2,0) -- (4,2) -- (4,4);
  \draw (0,0) -- (-1,-1);
  \draw (4,0) -- (5,-1);
  \draw (4,4) -- (5,5);
  \draw (0,4) -- (-1,5); 
  \draw (2,-2) node {(g)};
\end{scope}
\begin{scope}[shift={(10,-20)}]
  \draw (0,0) -- (4,0) -- (4,4) -- (0,4) -- (0,0);
  \draw (0,2) -- (4,2);
  \draw (2,0) -- (2,2-1.5*\gap);
  \draw (2,2+1.5*\gap) -- (2,4);
  \draw (0,0) -- (-1,-1);
  \draw (4,0) -- (5,-1);
  \draw (4,4) -- (5,5);
  \draw (0,4) -- (-1,5); 
  \draw (2,-2) node {(h)};
\end{scope}
\begin{scope}[shift={(20,-20)}]
  \draw (0,0) -- (2,4);
  \draw[color=white, line width=6pt] (1,3) -- (1.7,2.3);
  \draw (0,0) -- (4,0) -- (4,4) -- (0,4);
  \draw (0,4) -- (2,2) --  (4,2);
  \draw (2,0) -- (2,2);
  \draw (0,0) -- (-1,-1);
  \draw (4,0) -- (5,-1);
  \draw (4,4) -- (5,5);
  \draw (0,4) -- (-1,5); 
  \draw (2,-2) node {(i)};
\end{scope}
\end{tikzpicture}
\caption{The nine integral families needed to describe all master integrals for three-loop massless four-particle scattering.
The external legs are associated with the momenta $p_1$, $p_3$, $p_4$ and $p_2$ in clockwise order starting with the top left corner.}
\label{fig:3loopgraphs}
\end{figure}

In this paper we apply these methods to all three-loop integrals needed for two-to-two scattering.
The integrals can be arranged into nine integral families shown in Fig.~\ref{fig:3loopgraphs}.
The first analytical result for three-loop ladder boxes was obtained by one of the present authors in ref.~\cite{Smirnov:2003vi}.
The two planar families, (a) and (e) were computed previously in ref.~\cite{Henn:2013fah}.
Some of the non-planar integrals were computed in ref.~\cite{Henn:2013nsa}. 
In the present paper we report for the first time on the full set of integrals.

The paper is organized as follows. In section 2, we introduce our notations and conventions.
Then, in section 3, we present an improved version of the algorithm of ref.~\cite{WasserMSc} 
to find \dlogt integrands. 
In section 4, we explain how this can be combined with ideas from generalized unitarity, and 
point out differences. 
In section 5, we discuss practical aspects of the application of the algorithm, and 
comment on the scope of applications with the current implementation.
In section 6, we discuss the results of the application of the algorithm to three loops.
We also classify the resulting integrands according to their soft and collinear properties.
In section 7, we discuss the reduction to master integrals and the computation of
the latter using differential equations. We explain how we fix the boundary conditions
from physical consistency relations. Moreover, we discuss relations between integrals from
different integral families, and present a minimal set of master integrals.

\newpage

\section{Conventions, notation for integrands}
\label{sec:conventions}

In this section we introduce the notation and set-up for our computation of Feynman integrals contributing to four-particle scattering.
We denote the momenta of the four particles by $p_1\,\dots\,p_4$ and consider all of them to be in-going such that the momentum conservation identity
\beq
\label{eq:momcons}
p_1^\mu+p_2^\mu+p_3^\mu+p_4^\mu=0 
\eeq
is satisfied. The external particles we consider are massless and on-shell such that $p_i^2=0$. Furthermore, we define the Lorentz invariant scalar products 
\beq
s_{ij}=(p_i+p_j)^2 \,.
\eeq
Due to the specific kinematic scenario the following identity is satisfied:
\beq
\label{eq:Mandelrel}
s_{12}+s_{13}+s_{23}=0 \,.
\eeq
We always choose to eliminate the momentum $p_4$  using momentum conservation in our Feynman integrals. 
This in conjunction with the above equation allows us to express all our integrals in terms of only two variables $s$ and $t$.
We define
\beq
s=s_{12}\,,\hspace{1cm}t=s_{13},\hspace{1cm} x=-\frac{s_{13}}{s_{12}} \,.
\eeq
If we are describing a scattering process where particles with momenta $p_1$ and $p_2$ scatter and produce particles with momenta $p_3$ and $p_4$ then both $s$ and $x$ are positive and $x\in[0,1]$.

The Feynman integrals under consideration in this article are plagued by ultraviolet and infrared divergences which we regulate by working in the framework of dimensional regularization and using the generalized spacetime dimension 
\beq
D=D_0-2\epsilon \,.
\eeq
Above, $D_0$ is a generic even integer and can be specified to be $D_0=4$ in order to achieve physical results. 
Throughout this article we will denote Feynman integrals by the letter $J$ and differential forms that are integrated by the letters $\mathcal{I}$.
With this we may write
\begin{align}
J = \int \phi^{(D,L)} \mathcal{I}^{(D)} \,.
\end{align}
In the above equation we introduce furthermore a convenient normalization factor  that depends on the number of loops in the Feynman integral $L$.
\beq
\label{eq:normfactor}
\phi^{(D,L)}=\frac{e^{\gamma_{\rm E} \frac{D-D_0}{2}L}}{(i \pi^{D/2})^L}\,,
\eeq
where $\gamma_{\rm E}$ is the Euler-Mascheroni constant.

\section{Computing \dlogt forms algorithmically}
\label{sec:DlogAlgorithm}

Feynman integrands with integrands that can be written as \dlogt forms are important,
as they (conjecturally) evaluate to uniform weight functions after integration.
In this section, we discuss a systematic way of finding Feynman integrands with this property. 
We introduce the necessary concepts, and illustrate the individual steps of the algorithm
by examples.

\subsection{\dlogt forms and leading singularities}
\label{sec:dlogdef}

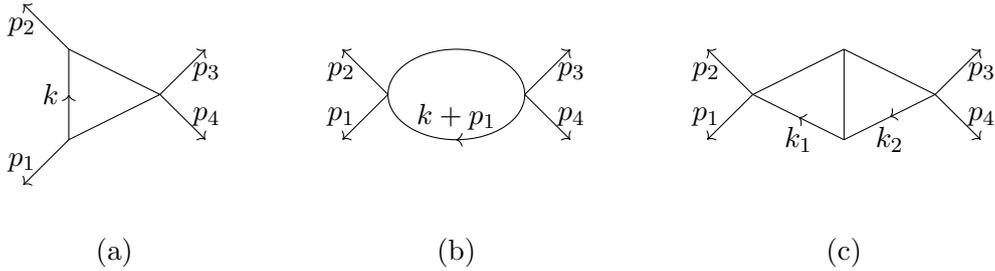
\begin{figure}[t]
\centering
\begin{tikzpicture}[scale=.6]
 \begin{scope}[shift={(-6,0)}]
 \draw[->] (1,2) -- (1,3) -- (3,2) -- (1,1) -- (1,2) node[left]{$k$};
 \draw[->] (1,3) -- (0,4) node[pos=.5, left]{$p_2$};
 \draw[->] (1,1) -- (0,0) node[pos=.5, left]{$p_1$};
 \draw[->] (3,2) -- (4,3) node[pos=.5,right]{$p_3$};
 \draw[->] (3,2) -- (4,1) node[pos=.5,right]{$p_4$};
 \draw (2,-1.5) node {(a)};
 \end{scope}
 \begin{scope}[shift={(1,0)}]
   \draw (2.5,2) ellipse (1.5 and 1);
 \draw[-{>[scale=1.0]}] (2.5,1) -- (2.49,1) node[pos=.0,above]{$k+p_1$};
 \draw[->] (1,2) -- (0,3) node[pos=.5,left]{$p_2$};
 \draw[->] (1,2) -- (0,1) node[pos=.5,left]{$p_1$};
 \draw[->] (4,2) -- (5,3) node[pos=.5,right]{$p_3$};
 \draw[->] (4,2) -- (5,1) node[pos=.5,right]{$p_4$};
 \draw (2.5,-1.5) node {(b)};
 \end{scope}
 \begin{scope}[scale=2., shift={(4,-0.5)}]
  \draw (1,1.5) -- (2,2);
  \draw (1.5,1.25) -- (1,1.5) node[pos=0.,below]{$k_1$};
  \draw[->] (2,1) -- (1.5,1.25) node[pos=.5,below]{};
  \draw (2,2) -- (2,1) node[pos=.5,left]{};
  \draw (2,2) -- (3,1.5) node[pos=.5,left]{};
  \draw (2.5,1.25) -- (2,1) node[pos=0.,below]{$k_2$};
  \draw[->] (3,1.5) -- (2.5,1.25)  node[pos=.5,below]{};

   \draw[->] (1,1.5) -- (0.5,2) node[pos=.5,left]{$p_2$};
 \draw[->] (1,1.5) -- (0.5,1) node[pos=.5,left]{$p_1$};
 \draw[->] (3,1.5) -- (3.5,2) node[pos=.5,right]{$p_3$};
 \draw[->] (3,1.5) -- (3.5,1) node[pos=.5,right]{$p_4$};
 \end{scope}
   \draw (12,-1.5) node {(c)};
\end{tikzpicture}
\caption{
The integrand of the triangle shown in (a) is an example of a \dlogt form.
The integrand of the Feynman integrals shown in (b) and (c) has a (hidden) double pole in four dimensions.
}
\label{fig:doublepole}
\end{figure}

We are interested in (Feynman) integrands that have the property that they can be written as a differential form that behaves as $dx/x$ in each variable near singularities. 
More precisely, given a set of integration variables $x_{i}$, for $i=1,\ldots, n$ (typically, the components of the loop momentum), and external variables $y_j$ (such as Mandelstam invariants and masses, for example),we define the differential
\begin{align}
d = \sum_{i=1}^{n} dx_{i} \frac{\partial}{\partial {x_{i}}}\,.
\end{align}
Then an integrand admitting a \dlogt form can be written as
\begin{align}
\label{eq:dlogdef}
  \mathcal{I} = \sum_k c_k \, \dlog \,g_1^{(k)} \wedge \dlog \,g_2^{(k)} \wedge ... \wedge \dlog \,g_n^{(k)} \,.
\end{align}
Here and in the following the wedge corresponds to the usual definition of a differential form giving rise to an oriented volume after integration, such that e.g. $dx_1 \wedge dx_2 = -dx_2 \wedge dx_1$.
We see that for each term in the sum, one could change variables from $x_{i}$ to the set $g_i^{(k)} =: \tau_{i}$.
The corresponding term would then look like
\begin{align}
c_{k} \, \dlog \, \tau_1 \wedge \ldots \wedge \dlog(\tau_n)  = c_{k} \, \frac{d \tau_1}{\tau_1} \wedge \ldots \wedge \frac{d \tau_n}{\tau_n}  \,.
\end{align}
Consequently, it is evident that all singularities of (\ref{eq:dlogdef}) locally (in an appropriate set of variables) behave as $dx/x$.
The following comments are in order:
\begin{itemize}
\item Often, one is interested in loop integrals in $D_{0}-2\eps$ dimensions, for some integer $D_{0}$. Below, we mostly consider
properties of the $D_{0}$-dimensional part of the integrand. 
This turns out to be sufficient for our purposes here. See \cite{Chicherin:2018old} for a refined analysis that allows to discriminate between integrands
that vanish at $D=4$.
\item When analyzing integrands one may change variables from the loop momentum to some other convenient variable.
For the question about a \dlogt form of the integrand to be well defined it is important to allow only algebraic changes of
variables.
\item Two {\it integrands} may lead to the same integrated function, but differ for example by a total derivative that integrates to zero.
For example, we will see that the triangle integral of Fig.~\ref{fig:doublepole}(c) has a \dlogt integrand, while the bubble integral of Fig.~\ref{fig:doublepole}(a) does not,
although the two integrals are equivalent after integration. 
\end{itemize}

This \dlogt property of the integrand is sometimes referred to as integrands having only logarithmic singularities, as opposed to double poles. 
We emphasize that for this terminology to be meaningful, it is important to distinguish between integrands and integrals. 

The coefficients $c_{k}$ can be computed, in principle, by taking multiple residues, for example by evaluating the integrand along the contour encircling the poles at $\tau_{i} = 0$. 
The coefficients $c_{k}$ are called {\it leading singularities}. (In some abuse of notation, sometimes the locations $\tau_{i}=0$ are also called leading singularities.)

Let us illustrate the \dlogt property with some examples, following \cite{Bern:2014kca}.
The four-dimensional integrand of the triangle integral is given by
\begin{equation}
\label{eq:trik}
  \mathcal{I}_{3}^{(4)} = \frac{d^4k}{(k+p_1)^2 k^2 (k-p_2)^2}.
\end{equation}
 It is convenient to parametrize the loop momentum using spinor variables, $p_i = \lambda_i \tilde{\lambda}_i$, 
\begin{equation}
 \label{eq:spinor-helicity-1}
 k=\alpha_1 p_1+\alpha_2 p_2+\alpha_3 \lambda_1\tilde{\lambda}_2  \frac{ \langle 23 \rangle}{\langle 13 \rangle}+ \alpha_4 \lambda_2\tilde{\lambda}_1  \frac{ \langle 13 \rangle}{\langle 23 \rangle}\,.
\end{equation}
The two complex vectors multiplying $\alpha_3$ and $\alpha_4$ are orthogonal to $p_1$ and $p_2$. 
Their normalization was chosen such that they have zero helicity weights. This implies that scalar 
products with other vectors can always be rewritten in terms of the standard Lorentz invariants $s$ and $t$.
This change of variables leads to  $d^{4}k \sim s^2 d\alpha_1\wedge d\alpha_2\wedge d\alpha_3 \wedge d\alpha_4$. 
(Here and in the following we tacitly drop numerical multiplicative factors.)
Plugging this into equation \eqref{eq:trik}, we obtain
\begin{align}
\label{eq:tri}
 \mathcal{I}_3^{(4)}=\frac{d\alpha_1\wedge d\alpha_2\wedge d\alpha_3 \wedge d\alpha_4}
 {s(\alpha_1 \alpha_2-\alpha_3 \alpha_4) (\alpha_1 \alpha_2-\alpha_3 \alpha_4+\alpha_2) (\alpha_1 \alpha_2-\alpha_3 \alpha_4-\alpha_1)}.
\end{align}
One may verify (by differentiation) that this can be rewritten in the following way
\begin{gather}
\label{eq:trildog}
 \mathcal{I}_3^{(4)}=\frac{1}{s}\dlog(\alpha_1 \alpha_2-\alpha_3 \alpha_4)\wedge \dlog(\alpha_1 \alpha_2-\alpha_3 \alpha_4+\alpha_2) \notag\\
 \wedge\dlog(\alpha_1 \alpha_2-\alpha_3 \alpha_4-\alpha_1)\wedge\dlog \alpha_3  \,.
\end{gather}
This is of the form of eq. (\ref{eq:dlogdef}). Remarkably, only a single term is needed.
We also see that the leading singularity of this diagram is ${1}/{s}$.\footnote{Note that leading singularities are only defined up to a numerical factor, since 
we can always rewrite dlog forms like
$ \dlog A = \frac{1}{2} \dlog A^2$.
} Of course, in this simple case this can also be seen by dimensional analysis.

One may make a further interesting observation. Written in momentum variables eq. (\ref{eq:trildog}) takes the form
\begin{align}\label{eq:trildog2}
 \mathcal{I}_3^{(4)} =  \frac{1}{s}\dlog k^2\wedge \dlog (k+p_1)^2\wedge\dlog(k-p_2)^2\wedge\dlog \, k \cdot k_{+}^* \,.
\end{align}
Here $k_{+}^*=\beta  \lambda_2\tilde{\lambda}_1$ , for arbitrary $\beta$.
(Obviously, (\ref{eq:trildog2}) is independent of $\beta$.) A similar formula holds with $k_{-}^*=\beta  \lambda_1\tilde{\lambda}_2$.

When the triangle integrand is written in the form (\ref{eq:trildog2}) we can see a close relationship between generalized unitarity and leading singularities.
It might appear surprising at first sight that one may take a four-fold residue for an integral having
only three propagators.
To see this, it is important to realize that  $k_{\pm}^{*}$ correspond to the two solutions of the maximal cut of the triangle integral.
The leading singularity $1/s$ can be computed by first taking the maximal cut, which corresponds to taking the residue at $k^2=0, (k+p_1)^2=0, (k-p_2)^2=0$.
Upon taking this maximal cut, a Jacobian factor is produced. For example, for one of the two possible cut solutions, this factor is $1/(s \alpha_3)$. 
So we have
\begin{align}
\oint_{(k-p_2)^2=0}  \oint_{(k+p_1)^2=0}  \oint_{k^2=0}  \mathcal{I}_3^{(4)} = \frac{1}{s} \, \frac{d \alpha_{3}}{\alpha_{3}} \,.
\end{align}
This form has new poles at $\alpha_3 = 0$ and $\alpha_3 =\infty$, which were not manifest in the original integrand (\ref{eq:tri}). 
The leading singularity $\pm 1/s$ is then obtained by taking a further residue at either of these poles.
Leading singularities involving such poles are called composite.

We remark that whenever a \dlogt representation of the form (\ref{eq:dlogdef}) is known, verifying it is relatively straightforward.
On the other hand, determining whether such a form exists for a given integrand, and computing it, is more complicated.
In the following we will present a method to derive \dlogt forms in an automated way.

Not all Feynman integrands admit the representation (\ref{eq:dlogdef}).
Whenever the integrand has a double or higher pole, 
it is impossible to rewrite it in the form $dx/x$ 
(restricting ourself to algebraic changes of variables).
For example, $\frac{d\alpha}{\alpha^2}$ does not admit a \dlogt form.
Similarly, if the integrand goes to a constant in some variable, this means that there is a double pole at infinity.

Note that double poles are not always obvious and sometimes are revealed after computing residues.
As an example, consider the bubble integral of Fig.~\ref{fig:doublepole}(b). Its integrand is
\begin{equation}
  \mathcal{I}_{2}^{(4)} = \frac{d^4k}{(k-p_{2})^2 (k+p_1)^2} \,.
\end{equation}
Using again the parametrization in eq. \eqref{eq:spinor-helicity-1} we have
\begin{align}
\label{eq:bubble}
  \mathcal{I}_{2}^{(4)}  =\frac{d\alpha_1\wedge d\alpha_2\wedge d\alpha_3 \wedge d\alpha_4}
 {[\alpha_1 (\alpha_2-1)-\alpha_3 \alpha_4] [(\alpha_1+1) \alpha_2  -\alpha_3 \alpha_4] }.
\end{align}
Taking residues at $\alpha_4={(\alpha_1 +1) \alpha_2}/{\alpha_3}$, then at $\alpha_3=0$, and
finally at $\alpha_2=-\alpha_1$, we find
\begin{equation}
\label{eq:bubbledb}
 \mathcal{I}_2^{(4),\, \text{cut}} = d\alpha_1 \,.
\end{equation}
We denoted the resulting form as a `cut' integrand (in analogy with generalized unitarity).
We see that the form in eq. (\ref{eq:bubbledb}) has a double pole at infinity, and hence $\mathcal{I}_{2}^{(4)}$ does not admit a \dlogt form. 
Note that this also implies that any multi-loop Feynman integrand 
with a bubble sub-loop cannot be written as a \dlogt form.

\subsection{Partial fractioning method}
\label{subsec:partialfrac}
In this section we show how partial fractioning can be used to systematically derive \dlogt forms and thereby also compute the leading singularities for a given integrand. 
The idea is very simple: we start with one integration variable (in principle, any), and partial fraction. 
We then write each fraction in that variable as the differential of a logarithm.
Then, we proceed with the next integration variable, and so on, until no further integration variables are left.

The question whether this algorithm terminates is closely related to the question whether the denominator is linearly reducible \cite{Panzer:2015ida}.
Making this property obvious may depends on a good parametrization of the given integrand.
For on-shell integrals, the type of spinor parametrization (\ref{eq:spinor-helicity-1}) turns out to be very useful.

Let us illustrate the method by reconsidering the massless triangle of the previous section.

After partial fractioning $\mathcal{I}_3^{(4)}$ in equation \eqref{eq:tri} with respect to $\alpha_1$, 
and writing the corresponding terms as differentials of logarithms, we have
\begin{align}
 \mathcal{I}_3^{(4)} = &-\frac{1}{\alpha _2 \alpha _3 \alpha _4 s}
 \dlog\left(\alpha _1 \alpha _2-\alpha _3 \alpha _4\right)  \nonumber \\
 &+\frac{1}{\alpha _2 \left(\alpha _2^2-\alpha _2+\alpha _3 \alpha _4\right) s}
 \dlog\left[ (\alpha _1+1) \alpha _2-\alpha _3 \alpha _4\right]  \nonumber \\
 &+\frac{\alpha _2-1}{\alpha _3 \alpha _4 \left(\alpha _2^2-\alpha _2+\alpha _3 \alpha _4\right) s}
 \dlog\left[ \alpha _1 (\alpha _2-1) -\alpha _3 \alpha _4 \right] \,.
\end{align}
Iterating this for the other integration variables we find the full integrand written as a sum of \dlogt forms:
\begin{align}
\label{eq:dlogtriangleexpanded}
 \mathcal{I}_3^{(4)} =& \frac{1}{s} \dlog\left(\alpha _4\right)\wedge \dlog\left(\alpha _2\right)\wedge \dlog\left(\alpha _3\right)\wedge \dlog\left(\alpha _2 \alpha _1-\alpha _1-\alpha _3 \alpha _4\right)\\
 &+\frac{1}{s} \dlog\left(\alpha _4\right)\wedge \dlog\left(\alpha _2\right)\wedge \dlog\left(\alpha _2^2-\alpha _2+\alpha _3 \alpha _4\right)\wedge \dlog\left(\alpha _1 \alpha _2+\alpha _2-\alpha _3 \alpha _4\right)\notag\\
 &-\frac{1}{s} \dlog\left(\alpha _4\right)\wedge \dlog\left(\alpha _2\right)\wedge \dlog\left(\alpha _3\right)\wedge \dlog\left(\alpha _3 \alpha _4-\alpha _1 \alpha _2\right)\notag\\
 &-\frac{1}{s} \dlog\left(\alpha _4\right)\wedge \dlog\left(\alpha _2\right)\wedge \dlog\left(\alpha _2^2-\alpha _2+\alpha _3 \alpha _4\right)\wedge \dlog\left(\alpha _2 \alpha _1-\alpha _1-\alpha _3 \alpha _4\right)\notag \,.
\end{align}
This is the direct output of the algorithm, and could be simplified. In particular, although it is not obvious, this representation is equivalent to eq. \eqref{eq:trildog}.
This illustrates the fact that \dlogt representations are not unique for given integrands. 

\subsection{Power counting constraints on numerators}
\label{subsec:numconstraint}
In this section we show how excluding double poles at infinity leads to certain power counting constraints.
This has an important application. It will allow us to write down, for a given (Feynman) denominator, 
a general numerator with a finite number of free parameters. The latter can then be fixed to find all possible 
\dlogt integrands for a given denominator.

As an example for the general idea, consider the following integrand
\begin{equation}
\label{eq:examplenum}
 \mathcal{I} = \frac{\mathcal{N}\,da \wedge db }{(a+s)b(a+b+s)} \,,
\end{equation}
where $s$ is an external variable.
We wish to construct the most general ansatz for a polynomial numerator that covers all possible \dlogt integrands for the given denominator.
One immediate observation we can make is that if the polynomial degree (in a given variable) of the numerator is equal or higher than that of the denominator,
there will be a double pole at infinity.
Using this constraint we conclude that the following ansatz is sufficient to cover all possible \dlogt forms for this denominator.
\begin{equation}
 \label{eq:exampleN}
 \mathcal{N}=n_1+n_2 a+n_3 b + n_4 a b \,.
\end{equation}
In principle, we could apply this simple power counting constraint directly 
to Feynman integrands, e.g. when written in the parametrization
 spinor variables (see eq. \eqref{eq:tri}).
However, for integrands built with propagators we can find even stronger constraints, as we explain presently.
Let us consider a general one-loop $n$-point integrand with loop momentum $k$ in integer dimension $D_{0}$, 
\begin{equation}
 I_{n,m}^{(D_{0})} =	\frac{d^{D_{0}}k \, N_{m}(k)}{k^2(k+p_1)^2(k+p_1+p_2)^2\cdots (k+p_1+...+p_{n-1})^2}.
\end{equation}
Here, we assume the numerator $N_{m}(k)$ to be a monomial of factors such as 
$k^2$ or  $k\cdot q_i$, with the total degree being $m$. 
Here $q_i$ being an arbitrary constant vector (e.g. an external momentum).

It turns out to be useful to perform a conformal inversion of the loop momentum \cite{Broadhurst:1993ib},
$k =  {\tilde{k}}/{\tilde{k}^2}$,
which implies
\begin{gather}
 d^4 k = \frac{d^{D_{0}}\tilde{k}}{(\tilde{k}^2)^{D_{0}}},\quad k^2 = \frac{1}{\tilde{k}^2}, \quad  k\cdot q = \frac{\tilde{k}\cdot q}{\tilde{k}^2} \,.
\end{gather}
This transformation reveals a double pole in $\tilde{k}^2$ for $n-m< D_{0}-1$.
Hence we find the constraint
\begin{equation}
\label{eq:powconstr}
n-m\ge D_{0}-1\,.
\end{equation}
Note that this is not the usual loop momentum power counting, since linear factors 
such as $k\cdot q$ and quadratic factors $(k+q)^2$ count the same.
For the triangle we then find that the only \dlogt numerator is a constant. 
We also find that the four-dimensional bubble integrand of eq. \eqref{eq:bubble} does not fulfill the power counting, 
which is consistent with having found a double pole.

Note also that the discussion so far was for a single term in the numerator. More generally, one can show that if $N$ is expanded in 
a basis of the monomials $k^2$ and $k\cdot p_i$, with $i=1,...,n-1$, the same power counting (\ref{eq:powconstr}) also applies to this situation, provided
that the basis terms are independent.

There is a subtlety related to the last point that we wish to address. 
Since we are performing the analysis in an integer dimension $D_{0}$, it 
is possible to write down linear combinations of terms that are equal to zero,
but in a non-trivial way. For example, consider the Gram determinant
{$G(k,p_1,p_2,p_3,p_4)$, with the loop momentum $k$ and four 
independent external momenta $p_1, ..., p_4$.} It vanishes
if the loop momentum is considered $D_{0}$-dimensional, but is non-zero
for $(D_{0}-2\eps)$-dimensional loop momentum.
Such linear combinations may contain terms that do
not fulfil the power counting constraint in equation \eqref{eq:powconstr}.
On the other hand, being zero, they are trivially \dlogt forms and therefore they seem to be a 
counterexample to the power counting criterium. Of course, this is not so, as the requirement
of independent basis terms was not met. 

In practice, it is desirable to control such evanescent Gram determinants in the numerator ansatz.
One may use the refined $D$-dimensional analysis of \cite{Chicherin:2018old}, where the integrand is written in a $D$-dimensional parametrization.
Using again the conformal transformation one can show that linear combinations which vanish in $D_{0}$ dimensions but violate the 
power counting constraint have double poles also in the $D$-dimensional analysis. 

The same power counting constraint can also be used for multi-loop integrands by applying the constraint loop by loop.

Empirically, we also found a more restrictive criterium at higher loops, namely
 \begin{equation}
 \label{eq:numconstraint}
n-m \ge \frac{D_{0}}{2}(L+1)-1 \,.
\end{equation}
While we do not necessarily expect this to be satisfied in general, we found it useful
as a restriction of the numerator ansatz at three loops. This point will be discussed further when
presenting the results.

\subsection{Dealing with non-linear denominator factors}
\label{sec:nonlin}
In the previous section we discussed how we are computing leading singularities by partial-fractioning denominators and subsequently by taking residues.
This procedure may be obstructed by denominators that are not linearly reducible.
In this subsection we describe how to proceed nevertheless in certain cases.
First, we discuss the case where at least one integration variable is at most quadratic in all denominator factors. 
Next, we discuss how to proceed in more general cases.

So we start with an integrand with a denominator
that is at most quadratic in all factors for 
some integration variable that we call $x$.
In a first step we make a partial fraction decomposition
with respect to $x$
such that all terms are either linear or quadratic in the
denominators. For the terms with linear numerators 
we can proceed in the standard way.
Terms with quadratic numerators have to be treated
differently and have the following general form:

\begin{equation}
\label{eq:quad}
 \frac{dx(u x+v)}{a x^2+b x+c},
\end{equation}
where $a, b, c, u$ and $v$ may depend on other integration variables.

There are two residues in $x$, which we denote by $r_1$ and $r_2$.
In other words, the integrand can be written as
\begin{equation}
 r_1 \dlog(x-s_1) + r_2 \dlog(x-s_2),
\end{equation}
where $s_1$ and $s_2$ are the two zeros of the quadratic denominator of equation \eqref{eq:quad}. 
Instead of processing with the computation with the residues of $r_1$ and $r_2$ in this form,
we can first simplify the expression. We do so by rewriting the last equation as
\begin{equation}
\label{eq:squaredlog}
 \frac{1}{2}(r_1+r_2)(\dlog(x-s_1)+\dlog(x-s_2))+\frac{1}{2}(r_1-r_2)(\dlog(x-s_1)-\dlog(x-s_2)),
\end{equation}
where 
\begin{align}
 r_1+r_2 &= \frac{u}{a},\\
 r_1-r_2 &= \frac{2av-ub}{a\sqrt{b^2-4ac}}.
\end{align}

Since $r_1+r_2$ is rational, for this term the computation again
can be continued with our standard methods.
The term $r_1-r_2$ has a square root in the denominator, so we have to find 
a way to deal with such a term.

In case the radicand is at most quadratic in one integration variable
$y$ and all other denominator factors are linear in $y$, we can proceed 
with the help of the following formulas:
\begin{gather}
 \frac{dy}{\sqrt{(y+a)(y+b)}} = 2 \dlog(\sqrt{y+a}+ \sqrt{y+b})\label{eq:sqrt1}, \nonumber \\
 \frac{dy}{y \sqrt{(y+a)(y+b)}} =  \frac{1}{\sqrt{a b}}\dlog \frac{y+\sqrt{y+a}\sqrt{y+b}-\sqrt{a}\sqrt{b}}{y+\sqrt{y+a}\sqrt{y+b}+\sqrt{a}\sqrt{b}}.\label{eq:sqrt2}
\end{gather} 
To apply these formulas we first have to 
do a partial fraction decomposition with respect to 
$y$ while treating the square root factor as a constant and possibly
do a constant shift in $y$ to get expressions of the form 
\eqref{eq:sqrt1} and \eqref{eq:sqrt2}.
Note that the residue in \eqref{eq:sqrt2}  is in general again a square root
of the remaining integration variables. So we can continue using the same formulas
for the next residue in case a suitable integration variable exists. It also may happen
that the residue is proportional to the square root of a perfect square. 
In this case the square root cancels and we may choose either sign of the square root.

Let us consider now two slightly more general cases.
Assume we have the following integrand
\begin{equation}
    \frac{dy\wedge dz N(y,z)}{(a y^2+b y+c z+d)\sqrt{P(y,z)}} \,,
\end{equation}
where $P(y,z)$ is a polynomial of degree at most two in $y$ and of degree higher than
two in $z$. Then neither $y$ nor $z$ fulfil the criteria for equation \eqref{eq:sqrt2} to
be applied. In this special case, however, we can make the following variable transformation in $z$:
\begin{equation}
    z \rightarrow \frac{b^2-4ad-4a^2z^2}{4ac} \,,
\end{equation}
which leads to
\begin{equation}
    a y^2+b y+c z+d \rightarrow \frac{(2 a y-2 a z+b) (2 a y+2 a z+b)}{4 a} \,.
\end{equation}
We see that the polynomial factorizes into two linear polynomials in $y$.
After this transformation the integrand has only linear factors in $y$ in the denominator
and the degree of $y$ in $P(y,z)$ does not change. This means that
we can do a partial fraction decomposition in $y$ and 
then apply equation $\eqref{eq:sqrt2}$.

The second special case is an integrand with only one integration variable:
\begin{equation}
    \frac{N(y) dy}{(ay^2+by+c)\sqrt{P(y)}}\,,
\end{equation}
where $P(y)$ is a polynomial of degree two or less in $y$. 
In this case we can force a factorization of $ay^2+by+c$ by also allowing square root 
terms of the external variables. After taking the residues we get a nested 
square root factor in the denominator, which does not cause a problem, because we 
do not take further residues. Often the radicand of the square root 
can be written as a perfect square and hence the 
nested square root can be simplified.

Finally we want to discuss the case of an integrand with a square root factor
in the denominator,
where the radicand polynomial is at least cubic in all integration variables.
In this case none of the methods discussed so far can be applied.
Here we try to proceed by performing a variable transformation depending on free parameters,
and then fix the latter in order to reduce the power degree for any of the integration variable in the radicand polynomial.

As an example for such a transformation, consider the following integrand
\begin{equation}
 \label{eq:trafoexample}
 \frac{dx\wedge dy}{(x+y) \sqrt{x^3+3 x^2 y+3 x^2+3 x y^2+2 x y+y^3}}.
\end{equation}
The polynomial of the square root is cubic in both variables $x$ and $y$,
so none of the methods discussed so far can be applied. 
So we make a parametrized variable transformation. For this example 
we consider the very simple type of transformation
\begin{equation}
 x \rightarrow x+\eta y.
\end{equation}
We find that for $\eta=-1$ the integrand simplifies to 
\begin{equation}
 \frac{dx\wedge dy}{x \sqrt{x^3+3 x^2-4 x y+y^2}},
\end{equation}
such that the radicand is now quadratic in $y$ and we can now take the residue
 in $y$ using \eqref{eq:sqrt1}.
For a general integrand with integration variables $z_1$ to $z_n$, we make the following transformations
\begin{equation}
\label{eq:trafoz}
 z_i \rightarrow z_i+\frac{Q(z_1,...,\hat{z}_i,...,z_n)}{(z_1 \cdots \hat{z}_i \cdots z_n)^{\nu}},
\end{equation}
with $i=1, ..., n$ and $\nu\in \{0,1\}$. Here $\hat{z}_i$ means that this variable is left out and $Q$ is 
a quadratic polynomial in all integration variables except $z_i$. We put a free coefficient $\eta_j$ before each term of the polynomial.
Since we transform only one variable at a time and $Q$ is independent of $z_i$ we do not 
change the integration measure with this transformation. After applying a transformation 
we check for each variable $z_h$, where $h=1,...,n$, if we can choose the free parameters $\eta_j$ such that all cubic and higher power terms of the radicand vanish. If a transformation 
of this type is found we apply equations \eqref{eq:sqrt1} or \eqref{eq:sqrt2} if the
requirements to the rest of the denominator are fulfilled. 
If we do not find a transformation the integrand remains unsolved.

\subsection{Algorithmic implementation}
\label{subsec:example}

\begin{figure}[t]
\centering
\begin{tikzpicture}[node distance=1.8cm]
\node (start) [startstop] {Rational integrand with ansatz for numerator};
\node (pro1) [process, below of=start] {1) Power counting to constrain ansatz};
\node (pro2) [process, below of=pro1] {2) Eliminate double poles};
\node (dec0) [decision, below of=pro2,  yshift=-1.5cm, text width=2.0cm] {3) Linear variable dependence?};
\node (pro4) [process, below of=dec0, yshift=-1.0cm] {4) Partial fraction and write form as sum of \dlogt's};
\node (pro5) [process, below of=pro4, text width=6cm] {5) Choose linearly independent set of the residues};
\node (dec1) [decision, below of=pro5, yshift=-1.0cm, text width=2cm] {6) Variables left?};
\node (pro6) [process, below of=dec1,  yshift=-1.0cm, text width=6cm] {7) Output: List of leading singularities and \dlogt form of the integrand};
\node (pro7) [process, below of=pro6] {8) Fix parameters to make leading singularities constant};
\node (stop) [startstop, below of=pro7] {Output: Complete \dlogt basis for given denominator structure};
\node (pro3) [process, right of=dec0, xshift=3.0cm, text width=4.4cm] {Variable transformation \\ Square root routines \\ (see section \ref{sec:nonlin})};

\draw [arrow] (start) -- (pro1);
\draw [arrow] (pro1) -- (pro2);
\draw [arrow] (pro2) -- (dec0);
\draw [arrow] (dec0) -- node[anchor=east]{yes}(pro4);
\draw [arrow] (pro4) -- (pro5);
\draw [arrow] (pro5) -- (dec1);
\draw [arrow] (dec1) -- node[anchor=east]{no}(pro6);
\draw [arrow] (pro6) -- (pro7);
\draw [arrow] (pro7) -- (stop);
\draw [arrow] (dec1) -- node[anchor=south]{yes}(2.5,-14.3) -- (9,-14.3) |- (pro2);
\draw [arrow] (pro2) -- (dec0);
\draw [arrow] (dec0) --  node[anchor=south]{no}(pro3);
\draw [dashed, arrow] (pro3) -- (8, -6.9) |- (0, -4.4);

\end{tikzpicture}
\caption{Workflow of the \dlogt algorithm}
 \label{fig:algonassi}
\end{figure}
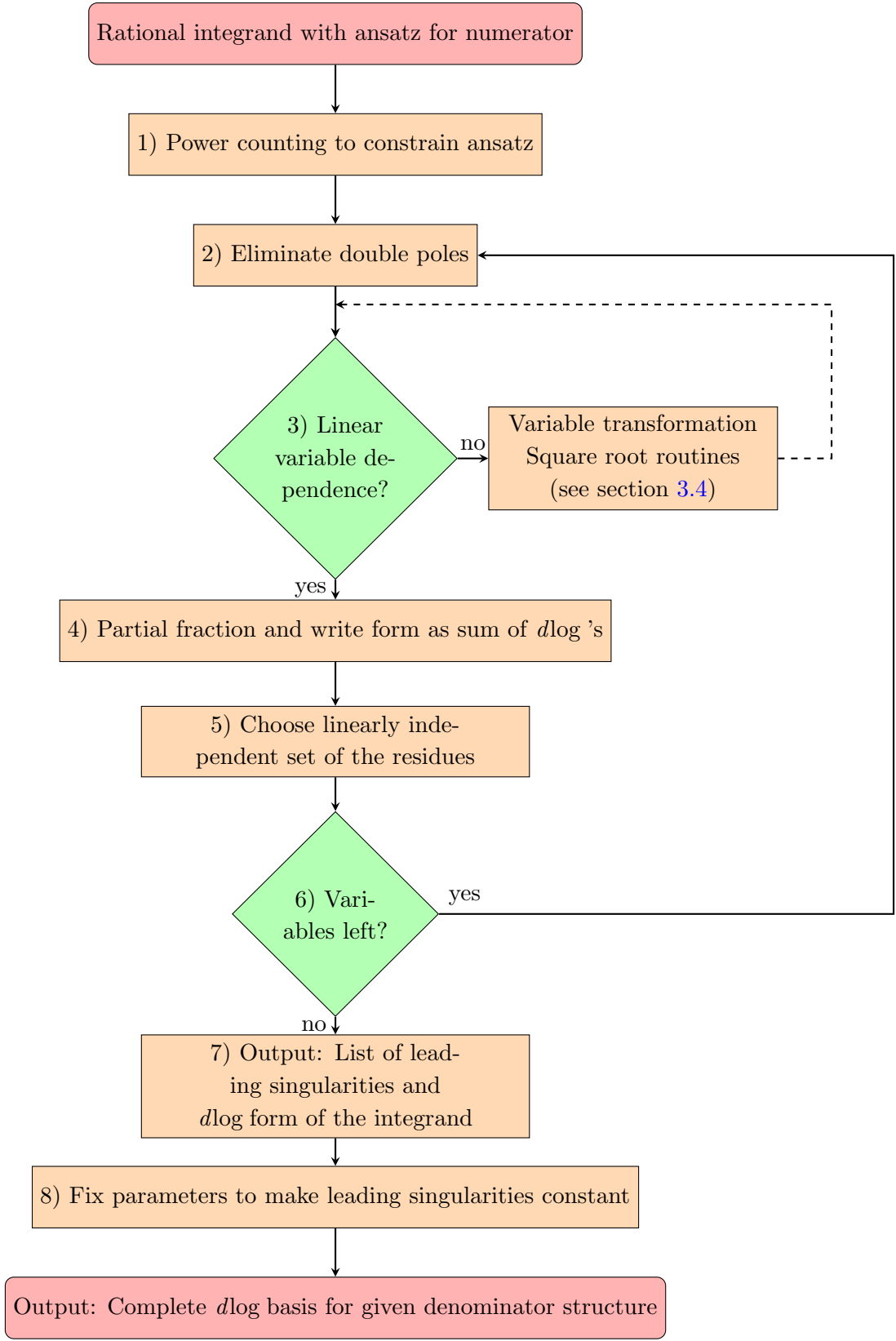

{The input is a denominator of an integrand, and the set of integration and external variables it depends on.
The denominator is required to be polynomial in the integration variables 
(one overall square root factor is also allowed).
The algorithm makes an ansatz for polynomial numerators.
It finds all numerators that have the property that the integrand 
can be written as a \dlogt form with constant leading singularities.
}

The algorithm, with all its steps, is visualized in Figure \ref{fig:algonassi}.
Let us go through them one by one, using the example of \eqref{eq:examplenum}.

{\underline{Step \#1}} consists in finding the most general numerator ansatz subject to power counting constraints,
as discussed in subsection \ref{subsec:numconstraint}.
In our example, the result of this step is given by equation \eqref{eq:exampleN}.

{\underline{Step \#2}} consists in eliminating double poles. 
Note that despite the initial constraints on the numerator, there might be further double poles 
in the integrand which get revealed by computing the leading singularities.
We will see an example of this below, in eq. (\ref{constraintn12}).

{\underline{Step \#3}}: We choose an integration variable that appears linearly in all denominator factors.
(If this is not possible, we continue with the method described in subsection \ref{sec:nonlin}.)
In the example, this is the case for both variables $a$ and $b$. Let us choose $b$.

{\underline{Step \#4}}: Partial fraction with respect to the variable chosen in step \#3, 
and write the terms as differentials of logarithms.
In our example, this yields
{
\begin{align}
  \mathcal{I} &= \frac{(n_1+a n_2)}{b(a+s)^2}da\wedge db+\frac{-n_1-a n_2+(a+s) n_3+a(a+s)n_4}{(a+s)^2(a+b+s)}da\wedge db\\
  & = \frac{n_1+a n_2}{(a+s)^2}da\wedge \dlog b+\frac{-n_1-a n_2+(a+s) n_3+a(a+s)n_4}{(a+s)^2}da \wedge \dlog(a+b+s) \notag.
\end{align}
}
Next, in {\underline{step \#5}} we find a linearly independent subset of the residues.
The residues are the factors multiplying the \dlogt factors. 
Choosing an independent set makes the subsequent calculation much more efficient.
In our example this step is trivial because there are only two residues that are obviously linearly independent.
In more complicated cases the list of residues is significantly longer.
The linear relations between the different residues can be found conveniently using numerical methods. 
For example, one may replace all external and internal variables by random integer numbers multiple times 
{(at least as many times as the number of residues)}
and then solve a system of linear equations. 

Having found the relations, we express all residues in terms of an independent basis,
and collect together all \dlogt terms having the same residue as a prefactor.
In practice, this step typically halves the number of terms {(which is typically of the same order as the the number of parameters in the numerator ansatz)}.
If we are interested in computing the leading singularities only, we may just keep
the independent residues, dropping the \dlogt factors.

In {\underline{step \#6}},  we check whether integration variables are left. If so, we continue with step \#2.
 So, in our example we again check for double poles. Indeed, at this stage there are factors 
$(a+s)^2$ in the denominator of both summands, indicating that the integrand 
has no \dlogt representation for generic $n_i$. We find the minimal
constraint on the free parameters $n_i$, such that the double pole vanishes. In other words, 
we demand the remainder of the polynomial division of the numerator and $(a+s)$ to vanish. This leads  
to the constraints
\begin{equation}\label{constraintn12}
 s \,n_2= n_1 \,,\quad  n_{4} = 0\,.
\end{equation}
Solving the constraints in eq. (\ref{constraintn12}) for the $n_{i}$, and proceeding with the next steps we obtain
\begin{align}
 \mathcal{I} &= \frac{n_1 da\wedge db}{s\,b(a+s)}+\frac{(-n_1+s n_3)da\wedge db}{s(a+s)(a+b+s)} \nonumber \\
 &= \frac{n_1 da}{s(a+s)}\wedge \dlog b + \frac{(-n_1+s n_3)}{s(a+s)}\wedge \dlog (a+b+s) \nonumber \\
 &= \frac{n_1}{s}\dlog(a+s)\wedge \dlog b+\left(-\frac{n_1}{s}+n_3 \right)\dlog(a+s)\wedge \dlog (a+b+s).
\end{align}

At this stage, there are no further integration variables left, so we proceed with {\underline{step \#7}}.
Here we identify the set of linearly independent leading singularities.
In our example, there are two of them, $n_1 /s $ and $-{n_1}/{s}+n_3 $.

Finally, in {\underline{step \#8}}, we find all solutions for the leading singularities to be constant numbers.
We do this in the following way. We take the list of $m$ linearly independent leading singularities
and solve the system of equations where one leading singularity is one and all others are zero. In this way we obtain 
$m$ independent solutions. In other words, this last step is just the inversion of a linear system of equations. 
Note that if this system has no solution it means that the numerator ansatz was incomplete.
On the other hand, if the solution depends on a parameter, this means that the numerator terms were not independent.

In our example, this is achieved e.g. by $(n_1, n_2 ) = (s,0)$ and $(n_1, n_2) = (0,1)$.
In other words, we find the following numerator solutions 
\begin{align}
 \mathcal{N}_1 &= a+s \,, \\
 \mathcal{N}_2 &= b \,.
\end{align}
This means we found a basis of all \dlogt forms with constant leading singularities for the given denominator.

{
Let us summarize the main steps. For a given denominator we write down 
a numerator ansatz that includes all possible \dlogt integrands, making use of power constraints. 
By repeatedly taking residues we reveal double poles that we exclude by constraining the parameters in the ansatz. 
After repeatedly taking residues, we eventually obtain a list of linearly independent leading singularities. 
We then find all solutions to the remaining parameters such that all leading singularities are constant numbers. 
In this way we construct, for the given denominator, a basis of integrands with a \dlogt form and constant leading singularities.
}

\FloatBarrier

\section{Cut-based organization of the calculation}
\label{sec:cuts}

\subsection{Similarities and differences to spanning set of cuts in unitarity approach}

Computing residues of Feynman integrands is obviously closely related to (generalized unitarity) cuts \cite{Bern:1994zx, Bern:1994cg}.
This is also very natural in the context of integration-by-parts (IBP)~\cite{Chetyrkin:1981qh,Tkachov:1981wb} relations and differential equations, as the matrices can be organized according
to integral sectors defined by cuts.
In particular, it is possible to organize the calculation into different parts by considering a so-called spanning 
set of cuts \cite{Larsen:2015ped}.
This has enormous potential, as it splits the calculation into smaller parts (parallelization), and moreover
each part is much simpler compared to the full calculation, and may be optimized further.

The spanning set of cuts in the context of IBP's corresponds to the maximal cuts of the master integrals that have no subsectors
with further master integrals. For the leading singularities we construct the spanning set of cuts in a very similar
way where instead of master integrals we consider all integrands that fulfil the power counting criterium defined in
section \ref{subsec:numconstraint}. 
This leads to a different notion of spanning cuts for computing all leading singularities to that in the context of IBP relations.
This can also be understood with the difference between four-dimensional integrands and integrands in $D$ dimensions.
As a consequence, for computing leading singularities in four dimensions one can 
in general take cuts with more propagators compared to the cuts that are used for IBP relations.

For example, in the context of $D$-dimensional IBP relations, the one-loop triangle integral of Fig.~\ref{fig:doublepole}(a)
is equivalent to the bubble integral of Fig.~\ref{fig:doublepole}(b), and hence to detect it one may cut 
the two propagators of the bubble only. On the other hand, in four dimensions there is no such relation, and the two
integrands are separate. In fact, the bubble integral is excluded by power counting. As a consequence, in this context it is 
sufficient to consider cuts with at least three propagators at one loop.

To compute a cut of propagators $P_1,...,P_n$, we solve
the equations $P_1=P_2=...=P_n=0$ for some integration variables $a_1, ..., a_n$
and then replace these propagators by the Jacobian 
$J=\det(\frac{\partial P_i}{\partial a_j})^{-1}$.
In this way we obtain an integrand where $n$ variables are already integrated out.
We then apply the \dlogt algorithm to the remaining integration variables 
In this way we obtain the leading singularities
and reveal double poles of the integrals on the cut.
Leading singularities on a cut are always a subset of the leading singularities of the whole integrand.
So the strategy is to combine all results of the different cuts until we have the complete list of leading singularities.

\subsection{Planar massless, on-shell double box in the cut-based approach}

We illustrate this method using the planar double box family as an example.
We follow the notation of \cite{WasserMSc}, where an early version of the \dlogt algorithm 
was used to analyze this family of integrals.

We define
\begin{align}
\label{eq:pldbint}
J_{a_1,...,a_9}=\frac{d^Dk_1d^Dk_2}{[-k_1^2]^{a_1}[-(k_1+p_1)^2]^{a_2}[-(k_1+p_1+p_2)^2]^{a_3}[-k_2^2]^{a_5}}\\
\times\frac{[-(k_1+p_1+p_2+p_3)^2]^{-a_4}[-(k_2+p_1)^2]^{-a_6}}{[-(k_2+p_1+p_2)^2]^{a_7}[-(k_1+p_1+p_2+p_3)^2]^{a_8}[-(k_1-k_2)^2]^{a_9}}\notag.
\end{align}
The ansatz for the numerator (subject to power counting) contains $26$ terms.
After eliminating double poles, $23$ \dlogt integrals are found, of which $10$ are
not related by flips of the graph that leave the kinematics invariant. These $10$ \dlogt integrals are shown in Table~\ref{tab:pldb}.
Using integration by parts (IBP) identities we find 8 master integrals which 
can be chosen from the 10 integrands. Since we have more \dlogt integrands than master integrals 
there are 2 IBP identities between integrals of the \dlogt basis. 
These IBP-relations are simple in the sense that they do not depend on external variables and the dimension, which can be explained by the 
fact that all \dlogt integrals have uniform transcendental weight. The two IBP relations are:
\begin{gather}
\label{eq:ibp1}
 j_1+j_2-j_3-\frac{1}{3}j_4-j_5=0,\\
\label{eq:ibp2}
 -4 j_2-\frac{14}{3} j_4-6 j_{5}+2j_{6}+j_7-j_8-3j_{9}+2j_{10} = 0.
\end{gather}

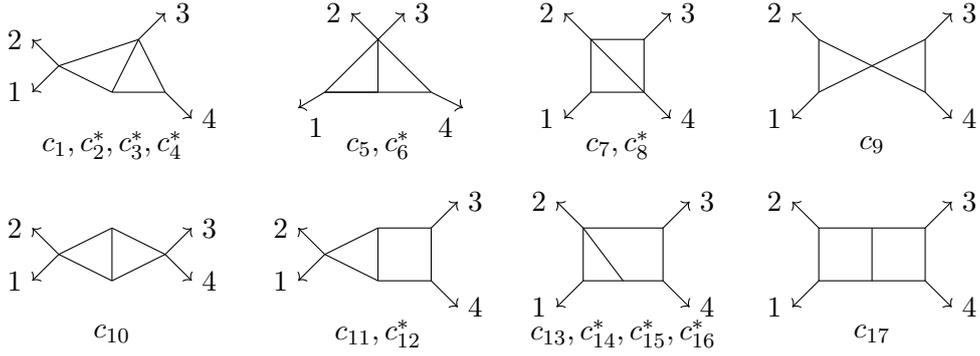
\begin{figure}
    \centering
    \begin{tikzpicture}
 \begin{scope}[shift={(0,0)},scale=0.7]
   \draw (1,1.5) -- (2,1);
   \draw (1,1.5) -- (2.5,2);
   \draw (2.5,2) -- (3,1);
   \draw (3,1) -- (2,1);
   \draw (2,1) -- (2.5,2);
   \draw[->] (1,1.5) -- (0.5,1.) node[left]{$1$}; 
   \draw[->] (1,1.5) -- (0.5,2.) node[left]{$2$};
   \draw[->] (2.5,2)-- (3,2.5) node[right]{$3$};
   \draw[->] (3,1) -- (3.5,0.5) node[right]{$4$};
  \draw (2,0) node {$c_1,c_2^*, c_3^*, c_4^*$};
 \end{scope}
 \begin{scope}[shift={(3.5,0)},scale=0.7]
\draw (1,1) -- (2,1);
\draw (1,1) -- (2,2);
\draw (2,2) -- (2,2);
\draw (2,2) -- (2,2);
\draw (2,2) -- (3,1);
\draw (3,1) -- (2,1);
\draw (2,1) -- (1,1);
\draw (2,1) -- (2,2);
\draw[->] (1,1) -- (0.5,0.7) node[below right]{$1$};
\draw[->] (2,2) -- (1.5,2.5) node[left]{$2$};
\draw[->] (2,2) -- (2.5,2.5) node[right]{$3$};
\draw[->] (3,1) -- (3.6,0.7) node[below left]{$4$};
  \draw (2,0) node {$c_5, c_6^*$};
 \end{scope}
 \begin{scope}[shift={(7,0)},scale=0.7]
   \draw (1,1) -- (2,1) -- (2,2) --(1,2) --(1,1);
   \draw (2,1) -- (1,2);
   \draw[->] (1,1) -- (0.5,0.5) node[left]{$1$};
   \draw[->] (1,2) -- (0.5,2.5) node[left]{$2$};<
   \draw[->] (2,2) -- (2.5,2.5) node[right]{$3$};
   \draw[->] (2,1) -- (2.5,0.5) node[right]{$4$};
  \draw (1.5,0) node {$c_7,c_8^*$};
 \end{scope}
  \begin{scope}[shift={(10,0)},scale=0.7]
   \draw (1,1) -- (2,1.5);
   \draw (1,1) -- (1,2);
   \draw (1,2) -- (2,1.5);
   \draw (2,1.5) -- (3,2);
   \draw (3,2) -- (3,1);
   \draw (3,1) -- (2,1.5);
   \draw (2,1.5) -- (2,1.5);
   \draw[->] (1,1) -- (0.5,0.5) node[left]{$1$};
   \draw[->] (1,2) -- (0.5,2.5) node[left]{$2$};
   \draw[->] (3,2) -- (3.5,2.5) node[right]{$3$};
   \draw[->] (3,1) -- (3.5,0.5) node[right]{$4$};
  \draw (2,0) node {$c_9$};
 \end{scope}
 \begin{scope}[shift={(0,-2.5)},scale=0.7]
   \draw (1,1.5) -- (2,1);
   \draw (1,1.5) -- (2,2);
   \draw (2,2) -- (3,1.5);
   \draw (3,1.5) -- (2,1);
   \draw (2,1) -- (2,2);
   \draw[->] (1,1.5) -- (0.5,1.) node[left]{$1$}; 
   \draw[->] (1,1.5) -- (0.5,2.) node[left]{$2$};
   \draw[->] (3,1.5)-- (3.5,2.) node[right]{$3$};
   \draw[->] (3,1.5) -- (3.5,1.) node[right]{$4$};
  \draw (2,0) node {$c_{10}$};
 \end{scope}
 \begin{scope}[shift={(3.5,-2.5)},scale=0.7]
   \draw (1,1.5) -- (2,1);
   \draw (1,1.5) -- (2,2);
   \draw (2,2) -- (3,2) -- (3,1) -- (2,1);
   \draw (2,1) -- (2,2);
   \draw[->] (1,1.5) -- (0.5,1.) node[left]{$1$}; 
   \draw[->] (1,1.5) -- (0.5,2.) node[left]{$2$};
   \draw[->] (3,2)-- (3.5,2.5) node[right]{$3$};
   \draw[->] (3,1) -- (3.5,.5) node[right]{$4$};
  \draw (2,0) node {$c_{11},c_{12}^*$};
 \end{scope}
\begin{scope}[shift={(7.25,-2.5)}, scale=.7]
  \draw (0.5,1) -- (2,1) -- (2,2) -- (0.5,2) -- (0.5,1);
  \draw (1.25,1) -- (0.5,2);
  \draw[->] (0.5,1) -- (0,0.5) node[left]{$1$}; ;
  \draw[->] (2,1) -- (2.5,0.5) node[right]{$4$};
  \draw[->] (2,2) -- (2.5,2.5) node[right]{$3$};
  \draw[->] (0.5,2) -- (0,2.5) node[left]{$2$};
  \draw (1.25,0) node {$c_{13},c_{14}^*,c_{15}^*,c_{16}^*$};
\end{scope}
\begin{scope}[shift={(10,-2.5)},scale=.7]
   \draw (1,1) -- (3,1) -- (3,2) -- (1,2) -- (1,1);
   \draw (2,1) -- (2,2);
   \draw[->] (1,1) -- (0.5,0.5) node[left]{$1$}; 
   \draw[->] (1,2) -- (0.5,2.5) node[left]{$2$};
   \draw[->] (3,2)-- (3.5,2.5) node[right]{$3$};
   \draw[->] (3,1) -- (3.5,0.5) node[right]{$4$};
  \draw (2,0) node {$c_{17}$};
\end{scope}
\end{tikzpicture}
    \caption{Sectors corresponding to the cuts used in the \dlogt analysis of
the planar double box family. 
Sectors corresponding to the three cuts in equation \eqref{eq:cuts} are $c_7, c_{13}, c_{17}$.
Labels with an asterisk represent sectors that can be obtained by flip symmetries and are not explicitly shown.}
    \label{fig:db_cuts}
\end{figure}

\setlength{\medmuskip}{0mu}
\begin{table}[t]
\centering
 \begin{tabular}{llll} 
  $j_{1} = s J_{1,0,1,0,1,0,0,1,1}, \;\;\;\;\;$
  &
  $j_{2} = t J_{1,1,0,0,1,0,0,1,1} $,
  & 
  $ j_{3} = (s+t) J_{1,1,0,0,0,0,1,1,1}$,
  \\
  $j_{4} = s t J_{1,1,0,0,1,0,1,1,1} $,
  &
  $j_{5} = s J_{1,1,0,0,1,-1,1,1,1}- s J_{1,1,0,0,0,0,1,1,1} $, 
  &
  $ j_{6} = s^2 J_{1,1,1,0,1,0,1,1,0} $,
  \\ 
  $j_{7} = s^2 J_{1,0,1,0,1,0,1,1,1}$ ,
  &
  $j_{8} = s^2 t J_{1,1,1,0,1,0,1,1,1}$,
  &
  $j_{9} = s^2 J_{1,1,1,0,1,-1,1,1,1} $,
   \end{tabular}
 \begin{tabular}{l}
  $j_{10} = s J_{1,1,1,-1,1,-1,1,1,1}-s J_{1,0,1,0,1,0,1,0,1}+s t J_{1,1,1,0,1,0,1,1,0}+t J_{0,1,1,0,1,0,0,1,1}+t J_{1,1,0,0,0,0,1,1,1}$.
 \end{tabular}
\caption{Planar double box integrands with constant leading singularities. 
Integrands that can be obtained from flip symmetries are not shown.
\label{tab:pldb}}
\end{table}
Let us now show how to derive these results in the cut-based approach.
When discussing cuts, let us use the following terminology.
If the propagators corresponding to a cut $c_A$ are a subset of the propagators
of a cut $c_B$, we say that $c_A$ is a subcut of $c_B$.
We find that for the double box family, given the numerator ansatz, there
is a total of $17$ cuts (see Figure \ref{fig:db_cuts}). In principle, we need to consider only $10$ cuts that do not have subcuts in that list, but to find \dlogt integrands of higher sectors more efficiently we construct the solution using
all cuts starting with the highest.

As an example let us consider the following three cuts
\begin{equation}
 \label{eq:cuts}
 \begin{aligned}
    c_7=\{1,1,0,0,0,0,1,1,1\}, \, 
    c_{13}= \{1,1,0,0,1,0,1,1,1\},\, 
    c_{17}= \{1,1,1,0,1,0,1,1,1\},
    \end{aligned}
\end{equation}
where the indices with value $1$ correspond to propagators that are cut.
The only integrand we have to consider for cut $c_7$ is 
$J_{1,1,0,0,0,0,1,1,1}$.
Setting the five propagators of $c_7$ to zero and solving the equations
with respect to five of the eight integration variables we find four solutions.
The latter can be understood as the four different helicity 
configurations that can be chosen when all five propagators are on-shell  
(see \cite{Carrasco:2011hw} for a review on this topic).
We proceed to compute the leading singularities for these four integrands and find that they are all proportional to $1/({s+t})$.
So we can normalize the integrand by $(s+t)$ to make the leading singularities constant on the cut.

Similarly we compute the leading singularities on the other cuts for the integrals
in the corresponding sectors. For the three examples we find the following integrals 
with constant leading singularities on the corresponding cuts:
\begin{align}
 c_7:\,\,& (s+t)J_{1,1,0,0,0,0,1,1,1}. \label{eq:cut1}\\
 c_{13}:\,\,& stJ_{1,1,0,0,1,0,1,1,1},\,s J_{1,1,0,0,1,-1,1,1,1}. \label{eq:cut2}\\
  c_{17}:\,\,& s^2 t J_{1,1,1,0,1,0,1,1,1},\,s^2 J_{1,1,1,0,1,-1,1,1,1},\,
   s^2 J_{1,1,1,-1,1,0,1,1,1},\,s J_{1,1,1,-1,1,-1,1,1,1} \label{eq:cut3}.
\end{align}
Since $c_7$ has no subcut in the spanning cuts, we know that $(s+t)J_{1,1,0,0,0,0,1,1,1}$
is already a \dlogt integrand with constant leading singularities. 
For the cuts $c_{13}$ and $c_{17}$ we have to take into account that there might be additional
leading singularities or double poles on subcuts.
To compensate the additional leading singularities and cancel out possible double poles on 
the subcuts we might have to add integrands from the corresponding subsectors.
Let us consider the second integral of \eqref{eq:cut2}.
Computing its leading singularities on the subcut $c_7$ we find an additional leading
singularity.
So we make an ansatz, where we add a linear combination of all 
\dlogt integrals from the corresponding subsector. In this case there is just one
\dlogt integrand, which means that we have the following ansatz:
\begin{equation}
\label{eq:cutansatz}
    s J_{1,1,0,0,1,-1,1,1,1}+n_1 (s+t)J_{1,1,0,0,0,0,1,1,1},
\end{equation}
Computing the leading singularities on $c_7$ we find that after setting
$n_1=-\frac{s}{s+t}$ all leading singularities are constant on $c_7$. 
Analyzing other subcuts we do not find further leading singularities, so that \eqref{eq:cutansatz}
is the complete \dlogt integral for $n_1=-\frac{s}{s+t}$.
The result agrees with the corresponding \dlogt integral in Table~\ref{tab:pldb}.

For the fourth integral in \eqref{eq:cut3}, there is one difference in the analysis: this time we also find 
a double pole on the subcut $c_{10}$: $\{1,0,1,0,1,0,1,0,1\}$.  Adding $-s J_{1,0,1,0,1,0,1,0,1}$, 
the double pole cancels out. Adding further integrands from subsectors to account 
for the additional leading singularities on the corresponding subcuts we find the following solution
\begin{equation}\nonumber
    s J_{1,1,1,-1,1,-1,1,1,1}-s J_{1,0,1,0,1,0,1,0,1}+s t J_{1,1,1,0,1,0,1,1,0}
  +t J_{0,1,1,0,1,0,0,1,1}+t J_{1,1,0,0,0,0,1,1,1} \,,
\end{equation}
which we also know already from Table~\ref{tab:pldb}.

\section{Practical comments and scope of applications}
\label{sec:improvements}

The basic version of the \dlogt algorithm was discussed in section \ref{sec:DlogAlgorithm}.
In addition, we use as an important further improvement the cut-based organization
of the calculation discussed in the last section.
Here we give further practical hints on the application to specific integrals, and
comment of the scope of the applications of the algorithm.

\subsection{Practical hints and comments}

In order to use the algorithm in a concrete application usually some preparatory steps need to be done.
We discuss these, as well as some hints for its efficient use.

\begin{itemize}

\item{\underline{ Parametrization of integration variables:}}
We find that for Feynman integrals with massless propagators, a 
spinor helicity parametrization such as eq. \eqref{eq:spinor-helicity-1} 
is quite efficient. As the latter involves the choice of two special on-shell momenta,
naturally, one may try different choices, as some may be better adapted to
a given diagram. (This is even more so when considering cuts.)
Let us mention also that a variant of the spinor helicity parametrization can also
be used in the case of massive external kinematics, by decomposing
a massive momentum in terms of two (arbitrary) light-like momenta.
Finally, we want to mention that another promising choice of paramterization
is the `improved Baikov' representation, see section 3.2 of \cite{Henn:2019swt}.

\item {\underline{Parametrization tailored to each cut:}} Choosing convenient parametrizations (of internal and external variables), and of the integration order, can 
be or practical importance. There is further potential for refinements in this direction in the cut-based approach: there, it may be natural
to choose a different parametrization tailored to each cut.

\item {\underline{Order of integration variables:}}
The algorithm analyzes a given integrand one integration variable at a time. After completing
the analysis in one variable, it may in principle proceed with any variable that fulfils that criteria
explained above. This gives a lot of possible orderings, and it may happen that the algorithm
terminates for some ordering, and not for other orderings. This is closely related to the question of linear reducibility \cite{Panzer:2015ida}.
Therefore running the algorithm
with different variable orderings may resolve some cases. This can naturally be parallelized.

\item {\underline{Dealing with square roots in the external kinematics:}}
Usually the external kinematics is expressed with a set of  Mandelstam invariants and masses.
In these variables, frequently square root factors appear in leading singularities . 
Sometimes it is possible to rationalize (some of) the square roots by changing the parametrization.
See e.g. \cite{Besier:2019kco} for an algorithmic implementation.
This can improve the performance of the algorithm, as it tends to minimize 
the number of square root terms encountered in intermediate steps.

\item {\underline{Special kinematics for problems with many variables:}}
Having many external variables may be another source of complications,
as this can make intermediate expressions grow easily to such an extent
that the computation is extremely slow or even not feasible.
In some cases, we already have a candidate integrand, and wish to test
whether it is a \dlogt form with constant leading singularities.
This can be particularly interesting with the method \cite{Hoschele:2014qsa,Dlapa:2020cwj} that requires only a single
UT integral to determine the complete UT basis.
In this case, we may e.g. replace all but one external variable by numerical constants and this way prove
for each variable individually that the leading singularity is independent of it.

\item {\underline{Integrands beyond integer dimensions:}} The computation of leading singularities is usually done for integrands with integer dimensions.
It turns out in most cases, integrands found from an analysis in integer dimensions can be straightforwardly upgraded to integrals with 
full dimensional dependence without losing the uniform transcendental weight property. 
Whenever this is not sufficient, a refined analysis is possible, as discussed in \cite{Chicherin:2018old}.
We find that for the integrals in the current paper this is not necessary.

\item {\underline{Simplified \dlogt forms:}} The output of the algorithm is a \dlogt form that can in principle be simplified further. 
Sometimes one can find representations with only a few or even a single term.
While this can be useful conceptually, and practically for direct integration \cite{Herrmann:2019upk}, this goes beyond the scope of this paper.

\end{itemize}

\subsection{Scope of applications}
The package provided with this paper was successfully applied to integrals with 
1) up to four loops,
2) up to five external variables,
3) integrals with massive propagators.
There are many examples where the computation can be done completely automatic using preimplemented
routines of the package only. In these cases we apply the following (standard) procedure:
\begin{itemize}
  \item Define kinematic setup.
  \item Use {\tt IntegrandAnsatz} to determine the set of integrands fulfilling the power counting constrains 
(see section \ref{subsec:numconstraint}). 
  \item For up to four external momenta from which one may be off-shell, we can directly use the 
routine {\tt SpinorHelicityParametrization} to define a parametrization and then use {\tt Parametrize} 
to parametrize the whole integrand ansatz. For other kinematic setups 
the parametrization must be set individually (see section \ref{sec:improvements}).
  \item Then use {\tt LeadingSingularities} to obtain all leading singularities and double pole constraints.
  \item Finally use {\tt GenerataeDlogbasis} to obtain the list of dlog integrands.
\end{itemize}
We will now discuss the scope of application considering different integral families and
discuss in which cases we used improvements to the standard procedure described above.

\begin{itemize}
  \item \underline{Three-loop four-point}: We computed \dlogt bases
of the three-loop four-point integral families (see Figure \ref{fig:3loopgraphs}). 
For all families except family (h) the computation can be done using the standard procedure.
Computing on a single kernel the computation time is between a few minutes for the simplest family (a) 
and 9 hours
for family (i) with up to 14 GB memory.
For the more complicated families (c), (f), (g), and (i) we used the package
together with  {\tt Macaulay 2} \cite{M2} to speed up the factorization of polynomials.
The \dlogt basis of Family (h) was obtained using the cut-based approach of section \ref{sec:cuts}.

\item \underline{Four loops}: 
An example for the successful application of the package to higher loop order are the four-loop
form factor integral families contributing to the quartic Casimir terms of
the light-like cusp-anomalous dimension in QCD \cite{Henn:2019rmi}.
Here again for the most complicated family (C) the cut-based approach is applied, while for
all other families the computation takes less than 16 hours on a single core each
using up to 2.2 GB memory with the standard procedure.

\item \underline{Massive propagators}: 
As a non-trivial example of integrals with massive propagators we apply the algorithm to 
two-loop integrals that appear e. g. in  $gg\rightarrow gg$ for a massive top
quark in the loop (see also \cite{Caron-Huot:2014lda}). 
In this case it is necessary to use a 
particular order of the integration variables. Finding a suitable order can be done by applying the
algorithm for different random variable orders (each run takes approximately two minutes) until
the computation is successful. The computation for this \dlogt basis is included in an example file.

\item \underline{Five-point two-loop integrals}: 
As an example for integrals with many scales we discuss 
the construction of \dlogt integrands for five-point two-loop integral families \cite{Chicherin:2018mue,Chicherin:2018old}.
Here additional steps to the routines implemented in the package are needed. 
While for the lower sectors the package can be used
with the standard procedure the \dlogt integrals of the higher sectors are more difficult to construct.
Due to Gram determinants 
in the integrand ansatz that vanish in the spinor helicity parametrization that was used 
throughout this paper, the leading singularities obtained this way are incomplete.
Hence, parts of the computations have to be performed for example in Baikov 
parametrization where these Gram determinants do not vanish.
Due to the many scales the \dlogt integrands are constructed using the cut-based approach and the external 
kinematic is chosen such that all leading singularities are rational functions.
\end{itemize}

\section{Results for \dlogt bases at three loops}
\label{subsec:3-loop}

\subsection{Description of method and results}

Here we apply our algorithm to compute \dlogt bases for the $9$ integral families shown in Fig.~\ref{fig:3loopgraphs}
(The labelling A to I follows \cite{Bern:2014kca}.) The classification of the planar families A and E was already done in \cite{WasserMSc}
and here we present the results for the non-planar families.

For the different integral families we again start with constructing a numerator ansatz,
subject to the power counting constraint discussed in section \ref{subsec:numconstraint},
including the heuristic one of eq. (\ref{eq:numconstraint}).
In this way, it turns out that the complication of Gram determinants is avoided, as the latter
would violate this condition.

We did the following consistency checks: 1) for the two planar families A and E, we checked that relaxing this constraint does
not lead to additional \dlogt solutions. 2) For all families, we checked that the ansätze are closed, in the following sense:
the number of independent leading singularities equals the number of free parameters.
A simple counterexample is the following list of leading singularities:
\begin{equation}
\left\{\frac{n_1}{s},\frac{n_1}{t} \right\} \,.
\end{equation}
Clearly, no choice of $n_1$ (except the trivial one) renders both leading singularities constant.
A complete ansatz is always closed. 
Therefore the fact that this does not happen supports the hypothesis that our ansatz did not miss \dlogt terms.

We found that for all families it was possible to chose a subset of \dlogt integrals as a basis of master integrals. In some cases it is necessary to consider
an integral family together with the same graph turned 90 degrees to get a complete \dlogt master integral basis.

The second column of Table~\ref{tab:3-loop} shows the size of the ansatz we used. The third column shows the number of \dlogt solutions for each integral family, 
where also symmetric equivalent solutions are counted. 
The fourth column counts the number of independent \dlogt integrals 
after applying integration by parts identities.
The fifth column gives the number of master integrals of the corresponding family. 

\begin{table}
\centering
\begin{tabular}{cccccc}	
Integral  & \# terms in & \# \dlogt forms  & \# independent & \# master integrals\\ 
family &numerator ansatz &found& \dlogt forms after IBP & in family \\
\hline
A & 141 & 101 & 25 & 26  \\
B & 215 & 168 & 47 & 47 \\
C & 307 & 205 & 50 & 53 \\
D & 325 & 256 & 28 & 28 \\
E & 281 & 171 & 41 & 41 \\
F & 325 & 199 & 62 & 62 \\
G & 377 & 253 & 87 & 87 \\
H & 651 & 440 & 76 & 76 \\
I & 451 & 325 & 113 & 113 
\end{tabular}
\caption{Application of the \dlogt algorithm to the 3-loop integral families.
\label{tab:3-loop}}
\end{table}

\subsection{Classification of the \dlogt integrals according to their infrared properties}

\begin{figure}[t]
\centering
\begin{tikzpicture}[scale=0.7]
 \begin{scope}[shift={(20,-20)}]
  \draw[->] (0,0) -- (2,0) node[pos=1, below]{$k_3$};
  \draw (2,0) -- (4,0);
  \draw (0,0) -- (0,2);
  \draw (0,2) -- (0,4);
  \draw (0,2) -- (4,4);
  \draw[color=white, line width=6pt] (1.6,2.8) -- (2.4,3.2);
  \draw[->] (0,4) -- (1,3.5) node[pos=1., above]{$k_1$};
  \draw (1,3.5) -- (4,2);
  \draw[->] (4,4) -- (4,3) node[pos=1., right]{$k_2$};
  \draw (4,3) -- (4,2);
  \draw (4,2) -- (4,0);
  \draw (4,0) -- (0,2);
  \draw[->] (0,0) -- (-1,-1) node[pos=0.5, above left]{$p_2$};
  \draw[->]  (4,0) -- (5,-1) node[pos=0.5, above right]{$p_4$};
  \draw[->]  (4,4) -- (5,5) node[pos=0.5, below right]{$p_3$};
  \draw[->]  (0,4) -- (-1,5) node[pos=0.5, below left]{$p_1$};
\end{scope}
\end{tikzpicture}
\caption{To reveal the $\epsilon^{-6}$ pole we take the following consecutive limits: 1) $\,k_1$ collinear to $p_1$, 2) $k_2$ collinear to $p_1$,
3) $k_2$ collinear to $p_3$, 4) $k_1$ collinear to $p_3$, 5) $k_3$ soft, 6) $k_3$ collinear to $p_2$.
The soft limit in $k_3$ contributes a pole only if applied after the series of four collinear limits in $k_1$ and $k_2$.}
\label{fig:diagrami}
\end{figure}
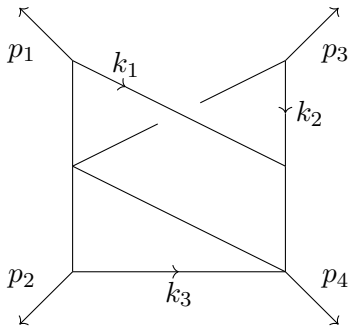

It turns out that all \dlogt integrals considered in this paper are ultraviolet finite,
thanks to the power counting constraints. 
So the only possible divergences after integration are of the soft/collinear type.
The latter are encoded into properties of the integrand, and are especially easy
to study for \dlogt integrals. 
It is therefore natural to classify the integrals in our basis according to their soft/collinear behavior,
following \cite{Henn:2019rmi} (cf. \cite{ArkaniHamed:2010gh,Drummond:2010mb,Bourjaily:2011hi} for earlier related work.)

To construct \dlogt integrals that are finite we take the linear combination of all {\dlogt} 
integrands and 
fix the coefficients such that the integrands vanish in all soft/collinear regions.
We investigate these regions by parametrizing loop momenta $k_i$ with a variable $x$ and consider the limit $x\rightarrow 0$.
For example we use $k_i = x \tilde{k_i}$ for a soft limit and
$k_i = \alpha p_1 + x^2 p_2 +x k_i^{\perp}$ for the limit where $k_i$ is collinear to $p_1$.
Applied to an integrand, we then have to cancel out factors such as $x^{-1- a \epsilon}dx$, for some $a$, which would result in a pole in $\epsilon$ after taking the integral
near $x = 0$.
Some infrared regions require multiple loop momenta being soft or collinear simultaneously.
Moreover, to reveal all poles in $\epsilon$ we find that in some cases it is necessary to consider consecutive $p_{j}$ and $p_{k}$ collinear limits of the same loop momentum $k_{i}$ as in eq. (6) of \cite{Henn:2019rmi}. For an example, see Fig.~\ref{fig:diagrami}.
We do the analysis for all possible momentum routings where $L$ propagators are written as $1/k_1^2, ..., 1/k_L^2$.

It is important to note that our construction corresponds to making the integrals finite {\it locally}.
This is different from integrals being finite due to some cancellation of $1/\eps^k$ poles
{\it after integration}, which is a weaker condition.
For example, classifying the $23$ \dlogt integrands of the planar double box, we find two finite integrals, in agreement with \cite{CaronHuot:2012ab}. 

In general we expect a given $L$-loop integral to have a pole of order $\epsilon^{-2L}$ at most.
To find integrals that are at most of order $\mathcal{O}(\epsilon^{-n})$ we
construct linear combinations of \dlogt integrals that vanish for any valid combination of $n+1$ 
infrared regions. For an $L$-loop integrand a valid combination can involve infrared limits of $L$ 
independent momenta. 
For each loop momentum we consider any pair of two external momenta and check for all soft/collinear regions. 
In doing so, we find it 
useful to employ a spinor parametrization based on those momenta.
In this way we can have poles of order $\epsilon^{-n}$ at most.

Since the number of different combinations of infrared regions quickly becomes very large ($\mathcal{O}(10^5)$ 
at three loops) we first apply them to the parent integrand of the integral family to sort out the combinations that
do not contribute. 
The remaining infrared limits can then be applied to the general linear combination of all
dlog integands in a parallelized computation.
Note that in order to have the complete list of combined limits we also have to consider infrared regions that
contribute a pole only after a certain combination of previous limits was applied. Figure \ref{fig:diagrami} shows
an example where the soft limit of $k_3$ contributes a pole only after a series of four collinear limits in the other 
loop momenta.

In this way, we classified all infrared poles of the integrals at three loops.
The results are shown in Table~\ref{tab:ir-3-loop}.
We provide the infrared ordered dlog integrals in an ancillary file to this paper.
\begin{table}[H]
\centering
\begin{tabular}{cccccccccc}	
 Family &  $\epsilon^0$ & $\epsilon^{-1}$  & $\epsilon^{-2}$  & $\epsilon^{-3}$  & $\epsilon^{-4}$  & $\epsilon^{-5}$  & $\epsilon^{-6}$\\
\hline
A & 8 & 16 & 18 & 24 & 19 & 0 & 16 \\
B & 0 & 32 & 34 & 36 & 28 & 0 & 38 \\
C & 22 & 24 & 36 & 48 & 31 & 0 & 44 \\
D & 0 & 0 & 96 & 48 & 36 & 0 & 76 \\
E & 10 & 36 & 36 & 32 & 35 & 0 & 22 \\
F & 8 & 15 & 45 & 42 & 50 & 0 & 39 \\
G & 10 & 41 & 47 & 53 & 46 & 0 & 56 \\
H & 0 & 70 & 98 & 88 & 56 & 0 & 128 \\
I & 0 & 48 & 79 & 56 & 66 & 0 & 76 \\
\end{tabular}
\caption{Number of \dlogt integrands with specific degree of divergence.}
\label{tab:ir-3-loop}
\end{table}

\section{All three-loop master integrals from differential equations}

In this section we discuss the analytic solutions of all 3-loop 4-point master integrals. 
First we define a set of 9 integral families that are sufficient to contain all required scalar Feynman integrals.
We label an integral of a family $\Lambda$ by
\beq
\label{eq:3LoopIntDef}
J_{\nu_1,\dots,\nu_{15}}^{\Lambda}=\int \phi^{(D,3)} d^Dp_5d^Dp_6d^D p_7\prod\limits_{i=1}^{15} \left(D_{\Lambda,i}^{-\nu_i}\right).
\eeq
The factor $\phi^{(D,3)}$ was defined in eq.~\eqref{eq:normfactor}.
We name the families by the first 9 letters in the alphabet, such that $\Lambda \in \{A,\dots,I\}$.
The factors $D_{\Lambda,i}$ correspond to integer linear combination of Lorentz invariant scalar products of external momenta and the loop momenta $p_5$, $p_6$ and $p_7$.
For example,
\beq
D_{A,2}=(p_1+p_2+p_5)^2.
\eeq
We define the $9\times 15$ factors $D_{\Lambda,i}$ in the ancillary files attached to the arXiv submission of this article. 
These factors are raised to generalized powers $\nu_i\in\mathbb{Z}$. 
However, the set of master integrals we are interested in satisfies $\nu_i\leq0$ for $i>10$.
The nine integrals $J^{\Lambda}_{1,1,1,1,1,1,1,1,1,1,0,0,0,0,0}$ corresponding to the nine integral families are represented graphically in fig.~\ref{fig:3loopgraphs}.

Next, we define a set of canonical master integrals $\vec{M}_\Lambda$ for each integral family. 
Any Feynman integral expressible in terms of the definition of eq.~\eqref{eq:3LoopIntDef} and with $\nu_i\leq0$ for $i>10$ can be related to our set of master integrals via IBP relations.
A master integral is a linear combination of Feynman integrals as defined in eq.~\eqref{eq:3LoopIntDef} with rational numbers and ratios of polynomials of Mandelstam invariants as pre-factors.
Additionally, we include a normalization factor $(D-4)^6 $ for each master integral. 
We find this canonical master integrals by applying the algorithm outlined in the previous sections.
In fact we find a complete basis for all families, except for one integral in family A and 3 integrals in family C using the algorithmic approach. 
For the missing four master integrals we select canonical integrals that have squared Feynman propagators. 
Such canonical integrals cannot be found by the algorithm in the form outlined above due to the power counting constraint. 
The number of required master integrals per family is presented in tab.~\ref{tab:3-loop}.
For example we choose,
\begin{eqnarray}
M_A^{25}&=&(D/2-2) ^6 \left(s_{12} \left(-s_{12}-s_{13}\right) J_{0,1,1,1,1,0,1,1,1,1,0,0,0,0,0}^A\right.\nonumber\\
&+&\left.s_{12}^3 J_{1,1,1,1,1,1,1,1,1,1,0,0,0,0,-1}^A+s_{12} \left(-s_{12}-s_{13}\right) J_{1,0,1,1,0,1,1,1,1,1,0,0,0,0,0}^A\right).
\end{eqnarray}
We give the definition of all chosen master integrals in the form of electronically readable files attached with the arXiv submission of this article.

In order to obtain a solution for our master integrals we apply the method of differential equations~\cite{Kotikov:1990kg,Kotikov:1991hm,Kotikov:1991pm,Henn:2013pwa,Gehrmann:1999as} in conjunction with IBP identities~\cite{Chetyrkin:1981qh,Tkachov:1981wb}. 
This allows us to write the total differential of our canonical master integrals in the form
\beq
\label{eq:equationDEX}
d\vec{M}_\Lambda=\epsilon\left[a\times \text{dlog} (s_{12})+b\times \text{dlog}(s_{13})+c \times \text{dlog}(s_{23})\right] \vec{M}_A.
\eeq
 $a$, $b$ and $c$ are matrices with rational rational entries. 
We emphasize that the canonical form of the differential equation~\eqref{eq:equationDEX} is obtained automatically since we are using \dlog~integrals as master integrals.
Next, we derive differential equations in the variable $x$ by applying the momentum conservation constraint of eq.~\eqref{eq:Mandelrel}.
Finally, we solve the resulting differential equations in an expansion in $\epsilon$ in terms of harmonic polylogarithms~\cite{Remiddi:1999ew} of argument $x$ up to $8^{\text{th}}$ order in the dimensional regulator. 

We determine the required boundary conditions from a simple physical requirement on how the solutions behave near singular points.
The matrices $a$, $b$ and $c$ have integer eigenvalues. 
We demand that the vector of our solutions evaluated at a singular point of the differential equations is in the kernel of the space spanned by the eigen-vectors correspoding to strictly positive eigen-values of the associated matrix $a$, $b$ or $c$.
The physical explanation of this constraint can be understood as follows. 
The solution of our differential equations to all orders in the dimensional regulator close to the point $s_{12}=0$ behaves as 
\beq
\lim\limits_{s_{12}\rightarrow 0} \vec{M}_\Lambda = s_{12}^{a\epsilon}  \vec{M}_{\Lambda,s_{12}=0}.
\eeq
Here, $ \vec{M}_{\Lambda,s_{12}=0}$ represents a vector of boundary constants. 
The matrix exponential $s_{12}^{\eps a}$ involves terms of the type $s_{12}^{\eps a_{i}}$, where $a_{i}$ are the eigenvalues of $a$ (in general positive and negative).
UV divergences are associated with infinitely small, but positive $\epsilon$.  
With the analysis of Feynman integrals we carried out in previous sections we demonstrated that it is possible to choose a basis of ultra-violet finite master integrals for any kinematical point, and in fact we did.
On the other hand, our solution to the differential equations would exhibit logarithmic UV divergences for positive $a_i$ at the point $s_{12}=0$ for a generic boundary condition.
To remedy this contradiction, we can choose the boundary vector $ \vec{M}_{\Lambda,s_{12}=0}$ such that no such divergences are present in our solution. 
The same has to be true for the other two singular points of our systems of differential equations, $s_{13}=0$ and $s_{23}=0$.
By expanding our general solution to the differential equations around all singular points and demanding this conditions have to be satisfied we constrain all boundary conditions except for the overall normalization.
We determine the latter by computing a trivial propagator type integral. 

We include our solutions to the differential equation as well as the systems of differential equations in ancillary files together with the arXiv submission of this article.
Our solution is valid in the scattering region outlined above, i.e. for $s_{12}>0$ and $x\in[0,1]$. 
Analytic continuation into another scattering region may be performed for example by following the steps discussed in ref.~\cite{Ahmed:2019qtg,Henn:2019rgj}. 
Similarly, permutations of external legs of our master integrals can be obtained using the methods detailed in refs.~\cite{Ahmed:2019qtg,Henn:2019rgj}. 
We checked that the permutations of our master integrals satisfy the permuted systems of differential equations.
Many master integrals that appear in one particular family also are contained within another and thus related to the master integrals of the other family.
We provide a complete set of master integrals for each family such that there are redundancies among our master integrals. 
In order to remove these redundancies we also provide in an ancillary file relations among master integrals across different families. 
These relations relate the total of 533 integrals as listed in Table~\ref{tab:3-loop} to 221 master integrals.

While not all Feynman integrals required for massless four-point scattering amplitudes at three loops can be expressed in terms of integrals in our families, we expect that all required master integrals can.
This expectation is based on the observation that all Feynman integrals that are not expressible in terms of our families contain sub-diagrams where at least one of the loops is in the form of a triangle integral.
Such integrals are always reducible via IBP identities and the resulting master integrals can be included within our nine integral families.

\section{Conclusion and future directions}

Above we outlined an algorithm to find Feynman integrals with a \dlog~integrand within a given integral family.
A preliminary version of the algorithm to find \dlog~forms  was presented in ref.~\cite{WasserMSc} and has already found multiple applications to cutting-edge problems.
These include four-loop non-planar form factor integrals \cite{Henn:2019rmi}, as well as two-loop integrals with many scales \cite{Chicherin:2018mue, Chicherin:2018old}. 
We discussed improvements of this algorithm and provide, for the first time, a public version. 

Efficient methods for obtaining \dlogt integrals are of great value in the process of computing Feynman amplitudes.
Such integrals allow to greatly facilitate the computation of Feynman integrals via the method of differential equations. 
Furthermore, we discussed connections of \dlog integrals and potential application within the framework for generalized unitarity. 
We also outlined how such integrals may be used to find Feynman integrals that are free of infrared and ultraviolet divergences or at least have lower degree of divergence.
 
Finally, we applied the algorithm to determine a basis of master integrals required to express any amplitude for the scattering of four massless particles at three loops with only massless virtual particles.
We then computed the master integrals using the method of differential equations.
Our solution takes the form of a  Laurent series in $\epsilon$ with coefficients that are expressed in terms of harmonic polylogarithms. 
In ancillary files attached to the arXiv submission of this article we provide a definition of these integrals, their explicit solution up to $\mathcal{O}(\epsilon^8)$ as well as the associated systems of differential equations.
With this, all integrals needed for virtual corrections to processes like di-jet or di-photon production at the LHC at next-to-next-to-next-to-leading order are known.

\acknowledgments

This research received funding from the European Research Council (ERC) under the European Union’s Horizon 2020 research and innovation programme,
{\it Novel structures in scattering amplitudes} (grant agreement No 725110).
B.M. is supported by a Pappalardo fellowship.
The work of V.S. was carried out according to the research program of Moscow Center of Fundamental and Applied Mathematics.
P.W. wants to thank Tiziano Peraro, Simone Zoia, Christoph Dlapa, Ekta Chaubey and Robin Brüser
for all the useful feedback on the implentation of the `DlogBasis' package.

\appendix
\section{Functions of DlogBasis package}
To use the {\tt DlogBasis} package the user needs {\tt Mathematica} to be installed (version 10 or higher).
The package can be downloaded using the command:\\ 
{\tt git clone https://github.com/pascalwasser/DlogBasis.git}

In the following we give an overview over the different functions implemented in the package.
The functions are also illustrated in an example file {\tt DlogBasis\_Examples.nb}.
\begin{itemize}
\item Load package:
\vspace{0.2cm}\\
\fbox{\parbox{0.91\textwidth}{\includegraphics[scale=0.91]{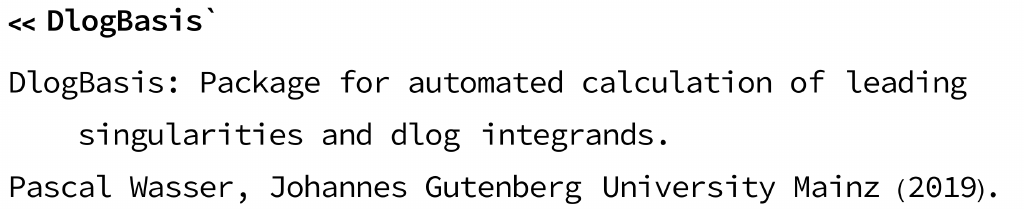}}
}\\
\item {\tt LeadingSingularities[func,v\_List],  LeadingSingularities[func,v\_List,n]}:\\ 
Computes the list of linear independent leading singularities for a given
multi-variable integrand. The integrand is either a rational function
or a rational function multiplied by a square root of a polynomial
in the denominator. The input is the integrand as the first argument
and the list of integration variables as the second argument. 
The output is the list of linear independent leading singularities.
If no dlog form exists
output is {\tt Fail[DoublePole]}.\vspace{0.1cm}\\
If the function 
is called with a third argument, {\tt func} must be a linear 
combination in the parameters {\tt n[1], ..., n[$m$]},
where {\tt n} is specified by the third argument. 
The output in this case is a list of two lists. The first is the list
of linear independent singularities. The second is a list of constraints
to the parameters to remove double poles.\vspace{0.1cm}\\
\fbox{\parbox{0.91\textwidth}{\includegraphics[scale=0.91]{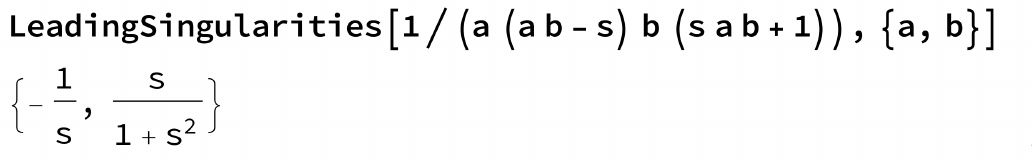}}
}\vspace{0.1cm}\\
\fbox{\parbox{0.91\textwidth}{\includegraphics[scale=0.91]{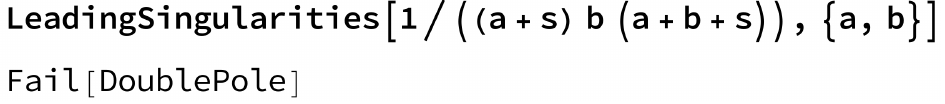}}
}\vspace{0.1cm}\\
\fbox{\parbox{0.91\textwidth}{\includegraphics[scale=0.91]{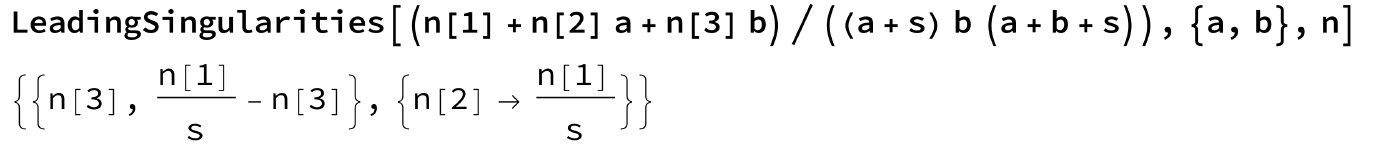}}
}\vspace{0.1cm}\\

\item {\tt InitializeDlogbasis[]}: \\
Initializes
the kinematic setup and is a necessary step for using the parametrization function
and to generate the integrand ansatz.
Note that {\tt LeadingSingularities} can also be called without any initialization.
The function is called after the
variables {\tt Internal} (loop momenta), {\tt External} (external momenta),
{\tt Replacements} (replacements of scalar products of external momenta) and 
{\tt Propagators} have been defined. The propagators can be
defined in terms of squared momenta (e. g. {\tt (k+p)\^{}2+mm}) and 
for linear propagators also in terms of scalar products (e. g. {\tt k.p})
written with the {\tt Dot}-symbol.
\vspace{0.1cm}\\
\fbox{\parbox{0.91\textwidth}{\includegraphics[scale=0.91]{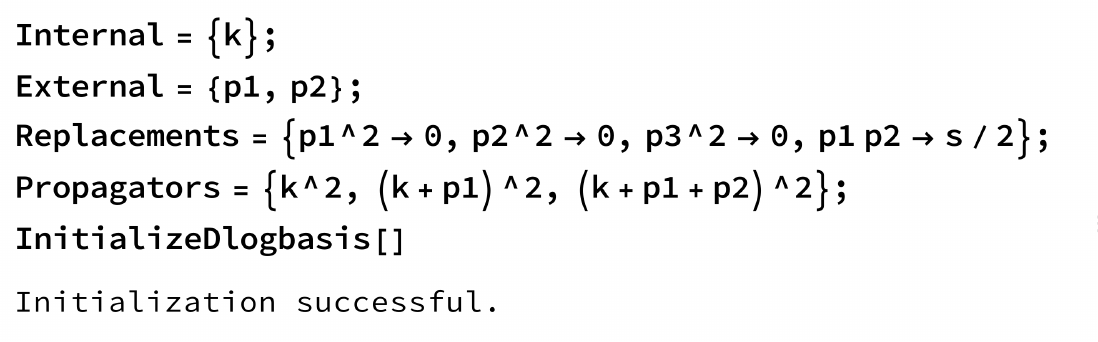}}
}\vspace{0.1cm}\\
\item {\tt SetParametrization[vs, eqs\_List, jac]}: \\
Initializes a parametrization of the loop momenta. 
The first argument is the list of new integration variables $v_{1},...,v_{4L}$.
The second argument defines the relation
between the original momentum variables and the new variables. If the 
equations do not parametrize all scalar products that 
depend on loop momenta, a warning is displayed.
The third argument is the jacobian $J$ of the coordinate transformation 
$d^4k_1\cdots d^4k_L = J dv_1 \cdots dv_{4L}$, where $k_1, ..., k_L$ are
the original loop momenta and $v_1, ..., v_{4L}$ are the new integration variables.\vspace{0.1cm}\\
\fbox{\parbox{0.91\textwidth}{\includegraphics[scale=0.91]{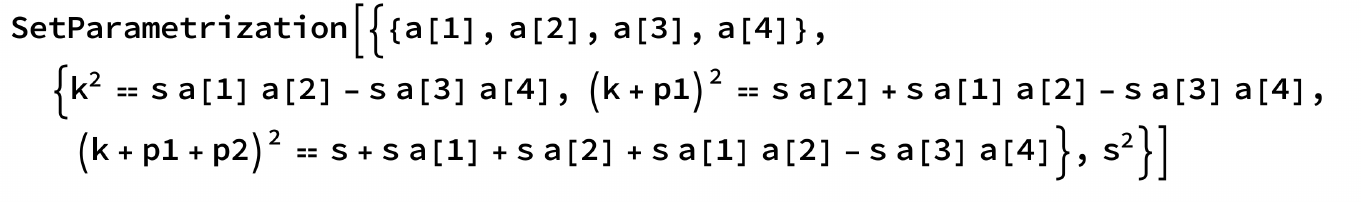}}
}\vspace{0.1cm}\\
\item {\tt SpinorHelicityParametrization[internal, vars, massless]}:\\
Generates a parametrization of the loop momenta and 
for massive external momenta in a spinor helicity basis (see equation
\eqref{eq:spinor-helicity-1}). 
The first argument is the list of internal (and possibly external) momenta
that should be parametrized. The second argument 
must have the same length as the first and defines the variable names 
for the parametrization variables. The last argument is a list of either 
two or three massless external momenta. The first two momenta define
the basis for the spinor helicity parametrization. The third momentum
is optional and defines a normalization factor to the mixed 
spinor vectors.
Output is a list with three elements. The first is the list of
integration variables, the second is the set of equations to 
define the scalar products and the third is the jacobian factor
for transforming the differential. The output can be directly 
used as an input for the {\tt SetParametrization} function.
\vspace{0.1cm}\\
\fbox{\parbox{0.91\textwidth}{\includegraphics[scale=0.91]{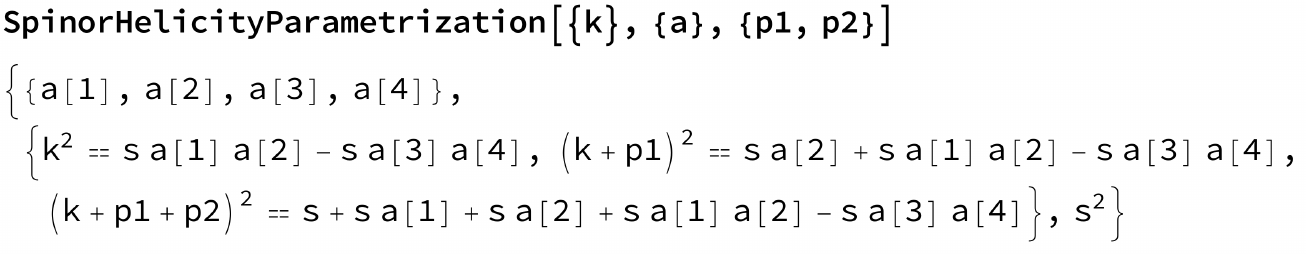}}
}\\
\fbox{\parbox{0.91\textwidth}{\includegraphics[scale=0.91]{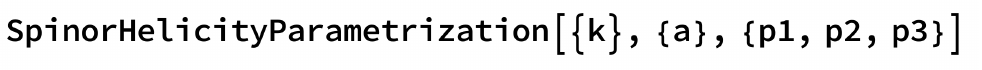}}
}\\
\item {\tt Parametrize[term], Parametrize[term\_List, n]}:\\
Parametrizes a given expression as specified with {\tt SetParametrization}.
Input is an arbitrary expression consisting of
scalar products, squared momenta and integrand terms
of the form {\tt G[fam, inds\_List]}.
Here {\tt fam}
is a label of the integral family that can be chosen by the user and
{\tt inds} is the list of propagator indices, which have to be integer numbers.
If a second argument {\tt n} is specified the first argument must 
be a list of terms {\{\tt t1,...,tm\}}. The output in this case is
the linear combination 
{\tt t1 n[1]+...+tm n[m]}, with {\tt t1,...,tm} parametrized.
\vspace{0.1cm}\\
\fbox{\parbox{0.91\textwidth}{\includegraphics[scale=0.91]{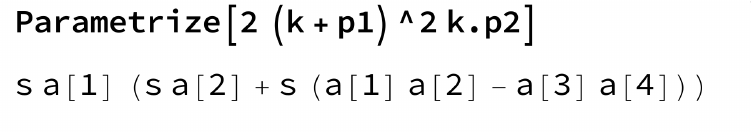}}
}\vspace{0.1cm}\\
\fbox{\parbox{0.91\textwidth}{\includegraphics[scale=0.91]{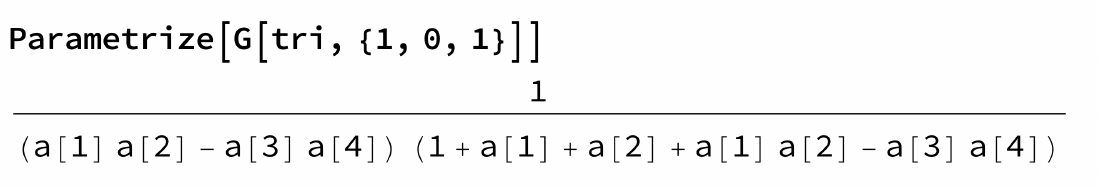}}
}\vspace{0.1cm}\\
\fbox{\parbox{0.91\textwidth}{\includegraphics[scale=0.91]{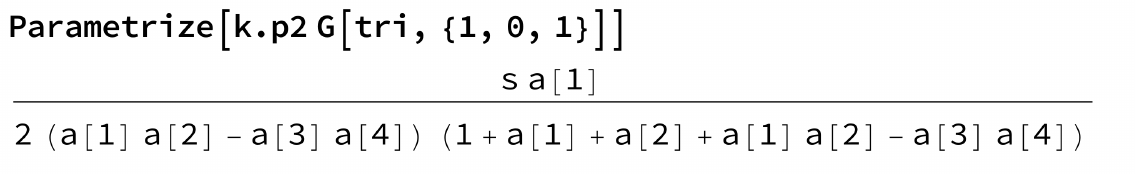}}
}\vspace{0.1cm}\\
\fbox{\parbox{0.91\textwidth}{\includegraphics[scale=0.91]{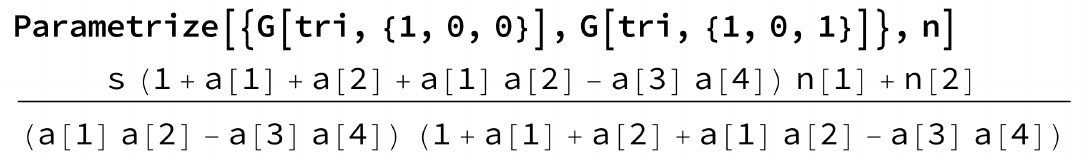}}
}\vspace{0.1cm}\\

\item {\tt IntegrandVariables[]}:\\Returns the list integration variables.\vspace{0.1cm}\\
\fbox{\parbox{0.91\textwidth}{\includegraphics[scale=0.91]{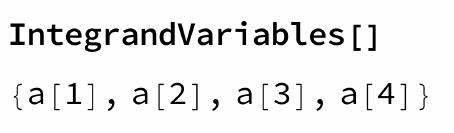}}
}\vspace{0.1cm}\\
\item {\tt IntegrandAnsatz[G[fam, inds\_List],dim:4]}:\\
The input is an integral without numerators in the form {\tt G[fam, inds\_List]}.
The list {\tt inds} must only contain values {\tt 1} and {\tt 0}.
For this integral all possible numerators are constructet,
which fulfill the \textit{dlog power counting} 
defined with equation \eqref{eq:numconstraint}. 
An optional second argument, which has to be an integer number, specifies the dimension and its default value is {\tt 4}.
The following example is the massless one-loop box family.
\vspace{0.1cm}\\
\fbox{\parbox{0.91\textwidth}{\includegraphics[scale=0.91]{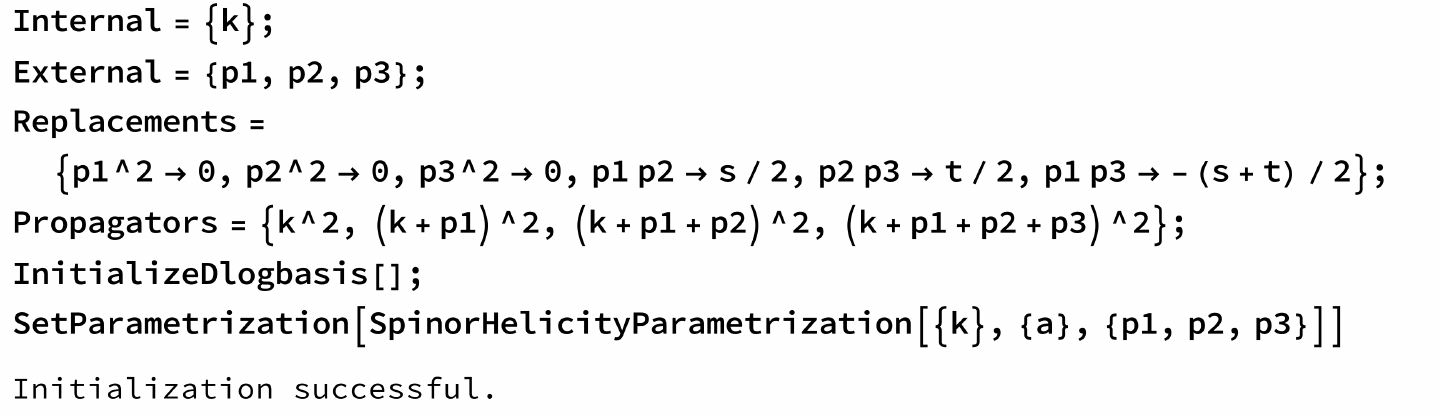}}
}\vspace{0.1cm}\\
\fbox{\parbox{0.91\textwidth}{\includegraphics[scale=0.91]{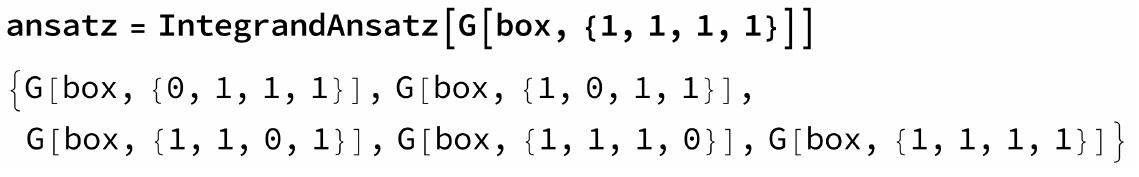}}
}\vspace{0.1cm}\\
\item {\tt GenerateDlogbasis[ansatz,lsing,n]}:\\
Converts a given integrand ansatz and list of leading singularities
into a list of dlog integrands with constant leading singularities.
The input are three arguments:
The first argument is the integrand ansatz. The second argument is the pair of leading
singularities and double pole constraints. The third argument 
is the variable name {\tt n} that defines the free
parameters {\tt n[1],n[2],...} of the leading singularities. 
If not all free parameters are fixed a warning is displayed.
The output is a list of dlog integrands
with constant leading singularities.\vspace{0.1cm}\\
\fbox{\parbox{0.91\textwidth}{\includegraphics[scale=0.91]{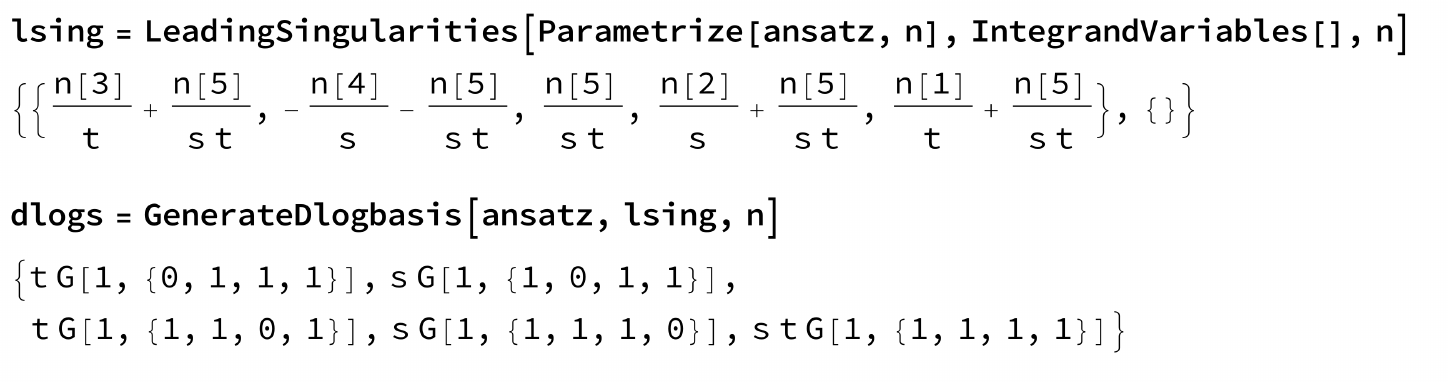}}
}\vspace{0.1cm}\\
\item {\tt UseMacaulay2[True/False]}:\\
Enables or disables the usage of Macaulay2 for a faster factorization of polynomials.
This function requires an installed version of Macaulay2. Furthermore the path to Macaulay2 must be assigned to the variable 
{\tt Macaulay2Path} and the variable {\tt DataPath} has to be set to a 
directory to save temporary files.
\vspace{0.1cm}\\
\fbox{\parbox{0.91\textwidth}{\includegraphics[scale=0.71]{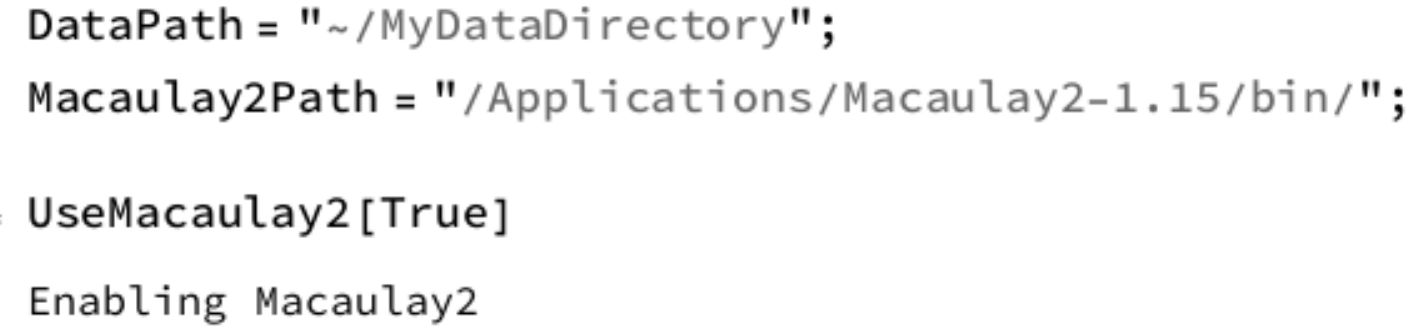}}
}\vspace{0.1cm}\\

\end{itemize}
\bibliography{BibFile}

\providecommand{\href}[2]{#2}\begingroup\raggedright\begin{thebibliography}{10}

\bibitem{Goncharov:1998kja}
A.~B. Goncharov, \emph{{Multiple polylogarithms, cyclotomy and modular
  complexes}}, \href{http://dx.doi.org/10.4310/MRL.1998.v5.n4.a7}{\emph{Math.
  Res. Lett.} {\bf 5} (1998) 497--516},
  [\href{https://arxiv.org/abs/1105.2076}{{\tt 1105.2076}}].

\bibitem{Chen:1977oja}
K.-T. Chen, \emph{{Iterated path integrals}},
  \href{http://dx.doi.org/10.1090/S0002-9904-1977-14320-6}{\emph{Bull. Am.
  Math. Soc.} {\bf 83} (1977) 831--879}.

\bibitem{Kotikov:2004er}
A.~V. Kotikov, L.~N. Lipatov, A.~I. Onishchenko and V.~N. Velizhanin,
  \emph{{Three loop universal anomalous dimension of the Wilson operators in
  $N=4$ SUSY Yang-Mills model}},
  \href{http://dx.doi.org/10.1016/j.physletb.2004.05.078,
  10.1016/j.physletb.2005.11.002}{\emph{Phys. Lett.} {\bf B595} (2004)
  521--529}, [\href{https://arxiv.org/abs/hep-th/0404092}{{\tt
  hep-th/0404092}}].

\bibitem{Henn:2016jdu}
J.~M. Henn and B.~Mistlberger, \emph{{Four-Gluon Scattering at Three Loops,
  Infrared Structure, and the Regge Limit}},
  \href{http://dx.doi.org/10.1103/PhysRevLett.117.171601}{\emph{Phys. Rev.
  Lett.} {\bf 117} (2016) 171601},
  [\href{https://arxiv.org/abs/1608.00850}{{\tt 1608.00850}}].

\bibitem{Abreu:2018aqd}
S.~Abreu, L.~J. Dixon, E.~Herrmann, B.~Page and M.~Zeng, \emph{{The two-loop
  five-point amplitude in $\mathcal{N} =4$ super-Yang-Mills theory}},
  \href{http://dx.doi.org/10.1103/PhysRevLett.122.121603}{\emph{Phys. Rev.
  Lett.} {\bf 122} (2019) 121603},
  [\href{https://arxiv.org/abs/1812.08941}{{\tt 1812.08941}}].

\bibitem{Caron-Huot:2019vjl}
S.~Caron-Huot, L.~J. Dixon, F.~Dulat, M.~von Hippel, A.~J. McLeod and
  G.~Papathanasiou, \emph{{Six-Gluon amplitudes in planar $ \mathcal{N} $ = 4
  super-Yang-Mills theory at six and seven loops}},
  \href{http://dx.doi.org/10.1007/JHEP08(2019)016}{\emph{JHEP} {\bf 08} (2019)
  016}, [\href{https://arxiv.org/abs/1903.10890}{{\tt 1903.10890}}].

\bibitem{Badger:2019djh}
S.~Badger, D.~Chicherin, T.~Gehrmann, G.~Heinrich, J.~M. Henn, T.~Peraro
  et~al., \emph{{Analytic form of the full two-loop five-gluon all-plus
  helicity amplitude}},
  \href{http://dx.doi.org/10.1103/PhysRevLett.123.071601}{\emph{Phys. Rev.
  Lett.} {\bf 123} (2019) 071601},
  [\href{https://arxiv.org/abs/1905.03733}{{\tt 1905.03733}}].

\bibitem{ArkaniHamed:2010gh}
N.~Arkani-Hamed, J.~L. Bourjaily, F.~Cachazo and J.~Trnka, \emph{{Local
  Integrals for Planar Scattering Amplitudes}},
  \href{http://dx.doi.org/10.1007/JHEP06(2012)125}{\emph{JHEP} {\bf 06} (2012)
  125}, [\href{https://arxiv.org/abs/1012.6032}{{\tt 1012.6032}}].

\bibitem{Gehrmann:2011xn}
T.~Gehrmann, J.~M. Henn and T.~Huber, \emph{{The three-loop form factor in N=4
  super Yang-Mills}},
  \href{http://dx.doi.org/10.1007/JHEP03(2012)101}{\emph{JHEP} {\bf 03} (2012)
  101}, [\href{https://arxiv.org/abs/1112.4524}{{\tt 1112.4524}}].

\bibitem{Drummond:2013nda}
J.~Drummond, C.~Duhr, B.~Eden, P.~Heslop, J.~Pennington and V.~A. Smirnov,
  \emph{{Leading singularities and off-shell conformal integrals}},
  \href{http://dx.doi.org/10.1007/JHEP08(2013)133}{\emph{JHEP} {\bf 08} (2013)
  133}, [\href{https://arxiv.org/abs/1303.6909}{{\tt 1303.6909}}].

\bibitem{Arkani-Hamed:2014via}
N.~Arkani-Hamed, J.~L. Bourjaily, F.~Cachazo and J.~Trnka, \emph{{Singularity
  Structure of Maximally Supersymmetric Scattering Amplitudes}},
  \href{http://dx.doi.org/10.1103/PhysRevLett.113.261603}{\emph{Phys. Rev.
  Lett.} {\bf 113} (2014) 261603}, [\href{https://arxiv.org/abs/1410.0354}{{\tt
  1410.0354}}].

\bibitem{Bern:2014kca}
Z.~Bern, E.~Herrmann, S.~Litsey, J.~Stankowicz and J.~Trnka, \emph{{Logarithmic
  Singularities and Maximally Supersymmetric Amplitudes}},
  \href{http://dx.doi.org/10.1007/JHEP06(2015)202}{\emph{JHEP} {\bf 06} (2015)
  202}, [\href{https://arxiv.org/abs/1412.8584}{{\tt 1412.8584}}].

\bibitem{WasserMSc}
P.~Wasser, \emph{{Analytic properties of Feynman integrals for scattering
  amplitudes}}, {\emph{M.Sc.} (2016) },
  [\href{https://arxiv.org/abs/https://publications.ub.uni-mainz.de/theses/frontdoor.php?source
  opus=100001967}{{\tt
  https://publications.ub.uni-mainz.de/theses/frontdoor.php?source
  opus=100001967}}].

\bibitem{Larsen:2017kzf}
K.~J. Larsen and R.~Rietkerk, \emph{{MultivariateResidues - a Mathematica
  package for computing multivariate residues}},
  \href{http://dx.doi.org/10.22323/1.290.0021}{\emph{PoS} {\bf RADCOR2017}
  (2017) 021}, [\href{https://arxiv.org/abs/1712.07050}{{\tt 1712.07050}}].

\bibitem{Kotikov:1990kg}
A.~V. Kotikov, \emph{{Differential equations method: New technique for massive
  Feynman diagrams calculation}},
  \href{http://dx.doi.org/10.1016/0370-2693(91)90413-K}{\emph{Phys. Lett.} {\bf
  B254} (1991) 158--164}.

\bibitem{Kotikov:1991hm}
A.~V. Kotikov, \emph{{Differential equations method: The Calculation of vertex
  type Feynman diagrams}},
  \href{http://dx.doi.org/10.1016/0370-2693(91)90834-D}{\emph{Phys. Lett.} {\bf
  B259} (1991) 314--322}.

\bibitem{Kotikov:1991pm}
A.~V. Kotikov, \emph{{Differential equation method: The Calculation of N point
  Feynman diagrams}}, \href{http://dx.doi.org/10.1016/0370-2693(91)90536-Y,
  10.1016/0370-2693(92)91582-T}{\emph{Phys. Lett.} {\bf B267} (1991) 123--127}.

\bibitem{Henn:2013pwa}
J.~M. Henn, \emph{{Multiloop integrals in dimensional regularization made
  simple}}, \href{http://dx.doi.org/10.1103/PhysRevLett.110.251601}{\emph{Phys.
  Rev. Lett.} {\bf 110} (2013) 251601},
  [\href{https://arxiv.org/abs/1304.1806}{{\tt 1304.1806}}].

\bibitem{Gehrmann:1999as}
T.~Gehrmann and E.~Remiddi, \emph{{Differential equations for two loop four
  point functions}},
  \href{http://dx.doi.org/10.1016/S0550-3213(00)00223-6}{\emph{Nucl. Phys.}
  {\bf B580} (2000) 485--518},
  [\href{https://arxiv.org/abs/hep-ph/9912329}{{\tt hep-ph/9912329}}].

\bibitem{Gituliar:2017vzm}
O.~Gituliar and V.~Magerya, \emph{{Fuchsia: a tool for reducing differential
  equations for Feynman master integrals to epsilon form}},
  \href{http://dx.doi.org/10.1016/j.cpc.2017.05.004}{\emph{Comput. Phys.
  Commun.} {\bf 219} (2017) 329--338},
  [\href{https://arxiv.org/abs/1701.04269}{{\tt 1701.04269}}].

\bibitem{Lee:2014ioa}
R.~N. Lee, \emph{{Reducing differential equations for multiloop master
  integrals}}, \href{http://dx.doi.org/10.1007/JHEP04(2015)108}{\emph{JHEP}
  {\bf 04} (2015) 108}, [\href{https://arxiv.org/abs/1411.0911}{{\tt
  1411.0911}}].

\bibitem{Meyer:2016slj}
C.~Meyer, \emph{{Transforming differential equations of multi-loop Feynman
  integrals into canonical form}},
  \href{http://dx.doi.org/10.1007/JHEP04(2017)006}{\emph{JHEP} {\bf 04} (2017)
  006}, [\href{https://arxiv.org/abs/1611.01087}{{\tt 1611.01087}}].

\bibitem{Prausa:2017ltv}
M.~Prausa, \emph{{epsilon: A tool to find a canonical basis of master
  integrals}}, \href{http://dx.doi.org/10.1016/j.cpc.2017.05.026}{\emph{Comput.
  Phys. Commun.} {\bf 219} (2017) 361--376},
  [\href{https://arxiv.org/abs/1701.00725}{{\tt 1701.00725}}].

\bibitem{Dlapa:2020cwj}
C.~Dlapa, J.~Henn and K.~Yan, \emph{{Deriving canonical differential equations
  for Feynman integrals from a single uniform weight integral}},
  \href{https://arxiv.org/abs/2002.02340}{{\tt 2002.02340}}.

\bibitem{Hoschele:2014qsa}
M.~Hschele, J.~Hoff and T.~Ueda, \emph{{Adequate bases of phase space master
  integrals for gg $\to$ h at NNLO and beyond}},
  \href{http://dx.doi.org/10.1007/JHEP09(2014)116}{\emph{JHEP} {\bf 09} (2014)
  116}, [\href{https://arxiv.org/abs/1407.4049}{{\tt 1407.4049}}].

\bibitem{Kosower:2011ty}
D.~A. Kosower and K.~J. Larsen, \emph{{Maximal Unitarity at Two Loops}},
  \href{http://dx.doi.org/10.1103/PhysRevD.85.045017}{\emph{Phys. Rev.} {\bf
  D85} (2012) 045017}, [\href{https://arxiv.org/abs/1108.1180}{{\tt
  1108.1180}}].

\bibitem{Badger:2012dv}
S.~Badger, H.~Frellesvig and Y.~Zhang, \emph{{An Integrand Reconstruction
  Method for Three-Loop Amplitudes}},
  \href{http://dx.doi.org/10.1007/JHEP08(2012)065}{\emph{JHEP} {\bf 08} (2012)
  065}, [\href{https://arxiv.org/abs/1207.2976}{{\tt 1207.2976}}].

\bibitem{Mastrolia:2012an}
P.~Mastrolia, E.~Mirabella, G.~Ossola and T.~Peraro, \emph{{Scattering
  Amplitudes from Multivariate Polynomial Division}},
  \href{http://dx.doi.org/10.1016/j.physletb.2012.09.053}{\emph{Phys. Lett.}
  {\bf B718} (2012) 173--177}, [\href{https://arxiv.org/abs/1205.7087}{{\tt
  1205.7087}}].

\bibitem{Ita:2015tya}
H.~Ita, \emph{{Two-loop Integrand Decomposition into Master Integrals and
  Surface Terms}},
  \href{http://dx.doi.org/10.1103/PhysRevD.94.116015}{\emph{Phys. Rev.} {\bf
  D94} (2016) 116015}, [\href{https://arxiv.org/abs/1510.05626}{{\tt
  1510.05626}}].

\bibitem{Bourjaily:2015jna}
J.~L. Bourjaily and J.~Trnka, \emph{{Local Integrand Representations of All
  Two-Loop Amplitudes in Planar SYM}},
  \href{http://dx.doi.org/10.1007/JHEP08(2015)119}{\emph{JHEP} {\bf 08} (2015)
  119}, [\href{https://arxiv.org/abs/1505.05886}{{\tt 1505.05886}}].

\bibitem{Bourjaily:2017wjl}
J.~L. Bourjaily, E.~Herrmann and J.~Trnka, \emph{{Prescriptive Unitarity}},
  \href{http://dx.doi.org/10.1007/JHEP06(2017)059}{\emph{JHEP} {\bf 06} (2017)
  059}, [\href{https://arxiv.org/abs/1704.05460}{{\tt 1704.05460}}].

\bibitem{Drummond:2010mb}
J.~M. Drummond and J.~M. Henn, \emph{{Simple loop integrals and amplitudes in
  N=4 SYM}}, \href{http://dx.doi.org/10.1007/JHEP05(2011)105}{\emph{JHEP} {\bf
  05} (2011) 105}, [\href{https://arxiv.org/abs/1008.2965}{{\tt 1008.2965}}].

\bibitem{Bourjaily:2011hi}
J.~L. Bourjaily, A.~DiRe, A.~Shaikh, M.~Spradlin and A.~Volovich, \emph{{The
  Soft-Collinear Bootstrap: N=4 Yang-Mills Amplitudes at Six and Seven Loops}},
  \href{http://dx.doi.org/10.1007/JHEP03(2012)032}{\emph{JHEP} {\bf 03} (2012)
  032}, [\href{https://arxiv.org/abs/1112.6432}{{\tt 1112.6432}}].

\bibitem{Henn:2019rmi}
J.~M. Henn, T.~Peraro, M.~Stahlhofen and P.~Wasser, \emph{{Matter dependence of
  the four-loop cusp anomalous dimension}},
  \href{http://dx.doi.org/10.1103/PhysRevLett.122.201602}{\emph{Phys. Rev.
  Lett.} {\bf 122} (2019) 201602},
  [\href{https://arxiv.org/abs/1901.03693}{{\tt 1901.03693}}].

\bibitem{vonManteuffel:2014qoa}
A.~von Manteuffel, E.~Panzer and R.~M. Schabinger, \emph{{A quasi-finite basis
  for multi-loop Feynman integrals}},
  \href{http://dx.doi.org/10.1007/JHEP02(2015)120}{\emph{JHEP} {\bf 02} (2015)
  120}, [\href{https://arxiv.org/abs/1411.7392}{{\tt 1411.7392}}].

\bibitem{vonManteuffel:2015gxa}
A.~von Manteuffel, E.~Panzer and R.~M. Schabinger, \emph{{On the Computation of
  Form Factors in Massless QCD with Finite Master Integrals}},
  \href{http://dx.doi.org/10.1103/PhysRevD.93.125014}{\emph{Phys. Rev.} {\bf
  D93} (2016) 125014}, [\href{https://arxiv.org/abs/1510.06758}{{\tt
  1510.06758}}].

\bibitem{Smirnov:2003vi}
V.~A. Smirnov, \emph{{Analytical result for dimensionally regularized massless
  on shell planar triple box}},
  \href{http://dx.doi.org/10.1016/S0370-2693(03)00895-5}{\emph{Phys. Lett.}
  {\bf B567} (2003) 193--199},
  [\href{https://arxiv.org/abs/hep-ph/0305142}{{\tt hep-ph/0305142}}].

\bibitem{Henn:2013fah}
J.~M. Henn, A.~V. Smirnov and V.~A. Smirnov, \emph{{Analytic results for planar
  three-loop four-point integrals from a Knizhnik-Zamolodchikov equation}},
  \href{http://dx.doi.org/10.1007/JHEP07(2013)128}{\emph{JHEP} {\bf 07} (2013)
  128}, [\href{https://arxiv.org/abs/1306.2799}{{\tt 1306.2799}}].

\bibitem{Henn:2013nsa}
J.~M. Henn, A.~V. Smirnov and V.~A. Smirnov, \emph{{Evaluating single-scale
  and/or non-planar diagrams by differential equations}},
  \href{http://dx.doi.org/10.1007/JHEP03(2014)088}{\emph{JHEP} {\bf 03} (2014)
  088}, [\href{https://arxiv.org/abs/1312.2588}{{\tt 1312.2588}}].

\bibitem{Chicherin:2018old}
D.~Chicherin, T.~Gehrmann, J.~M. Henn, P.~Wasser, Y.~Zhang and S.~Zoia,
  \emph{{All Master Integrals for Three-Jet Production at
  Next-to-Next-to-Leading Order}},
  \href{http://dx.doi.org/10.1103/PhysRevLett.123.041603}{\emph{Phys. Rev.
  Lett.} {\bf 123} (2019) 041603},
  [\href{https://arxiv.org/abs/1812.11160}{{\tt 1812.11160}}].

\bibitem{Panzer:2015ida}
E.~Panzer, \emph{{Feynman integrals and hyperlogarithms}}.
\newblock PhD thesis, Humboldt U., Berlin, Inst. Math., 2015.
\newblock \href{https://arxiv.org/abs/1506.07243}{{\tt 1506.07243}}.

\bibitem{Broadhurst:1993ib}
D.~J. Broadhurst, \emph{{Summation of an infinite series of ladder diagrams}},
  \href{http://dx.doi.org/10.1016/0370-2693(93)90202-S}{\emph{Phys. Lett.} {\bf
  B307} (1993) 132--139}.

\bibitem{Bern:1994zx}
Z.~Bern, L.~J. Dixon, D.~C. Dunbar and D.~A. Kosower, \emph{{One loop n point
  gauge theory amplitudes, unitarity and collinear limits}},
  \href{http://dx.doi.org/10.1016/0550-3213(94)90179-1}{\emph{Nucl. Phys.} {\bf
  B425} (1994) 217--260}, [\href{https://arxiv.org/abs/hep-ph/9403226}{{\tt
  hep-ph/9403226}}].

\bibitem{Bern:1994cg}
Z.~Bern, L.~J. Dixon, D.~C. Dunbar and D.~A. Kosower, \emph{{Fusing gauge
  theory tree amplitudes into loop amplitudes}},
  \href{http://dx.doi.org/10.1016/0550-3213(94)00488-Z}{\emph{Nucl. Phys.} {\bf
  B435} (1995) 59--101}, [\href{https://arxiv.org/abs/hep-ph/9409265}{{\tt
  hep-ph/9409265}}].

\bibitem{Chetyrkin:1981qh}
K.~G. Chetyrkin and F.~V. Tkachov, \emph{{Integration by Parts: The Algorithm
  to Calculate beta Functions in 4 Loops}},
  \href{http://dx.doi.org/10.1016/0550-3213(81)90199-1}{\emph{Nucl. Phys.} {\bf
  B192} (1981) 159--204}.

\bibitem{Tkachov:1981wb}
F.~V. Tkachov, \emph{{A Theorem on Analytical Calculability of Four Loop
  Renormalization Group Functions}},
  \href{http://dx.doi.org/10.1016/0370-2693(81)90288-4}{\emph{Phys. Lett.} {\bf
  100B} (1981) 65--68}.

\bibitem{Larsen:2015ped}
K.~J. Larsen and Y.~Zhang, \emph{{Integration-by-parts reductions from
  unitarity cuts and algebraic geometry}},
  \href{http://dx.doi.org/10.1103/PhysRevD.93.041701}{\emph{Phys. Rev.} {\bf
  D93} (2016) 041701}, [\href{https://arxiv.org/abs/1511.01071}{{\tt
  1511.01071}}].

\bibitem{Carrasco:2011hw}
J.~J.~M. Carrasco and H.~Johansson, \emph{{Generic multiloop methods and
  application to N=4 super-Yang-Mills}},
  \href{http://dx.doi.org/10.1088/1751-8113/44/45/454004}{\emph{J. Phys.} {\bf
  A44} (2011) 454004}, [\href{https://arxiv.org/abs/1103.3298}{{\tt
  1103.3298}}].

\bibitem{Henn:2019swt}
J.~M. Henn, G.~P. Korchemsky and B.~Mistlberger, \emph{{The full four-loop cusp
  anomalous dimension in $\mathcal{N}=4$ super Yang-Mills and QCD}},
  \href{https://arxiv.org/abs/1911.10174}{{\tt 1911.10174}}.

\bibitem{Besier:2019kco}
M.~Besier, P.~Wasser and S.~Weinzierl, \emph{{RationalizeRoots: Software
  Package for the Rationalization of Square Roots}},
  \href{https://arxiv.org/abs/1910.13251}{{\tt 1910.13251}}.

\bibitem{Herrmann:2019upk}
E.~Herrmann and J.~Parra-Martinez, \emph{{Logarithmic forms and differential
  equations for Feynman integrals}},
  \href{https://arxiv.org/abs/1909.04777}{{\tt 1909.04777}}.

\bibitem{M2}
D.~R. Grayson and M.~E. Stillman, ``Macaulay2, a software system for research
  in algebraic geometry.'' Available at
  \url{http://www.math.uiuc.edu/Macaulay2/}.

\bibitem{Caron-Huot:2014lda}
S.~Caron-Huot and J.~M. Henn, \emph{{Iterative structure of finite loop
  integrals}}, \href{http://dx.doi.org/10.1007/JHEP06(2014)114}{\emph{JHEP}
  {\bf 06} (2014) 114}, [\href{https://arxiv.org/abs/1404.2922}{{\tt
  1404.2922}}].

\bibitem{Chicherin:2018mue}
D.~Chicherin, T.~Gehrmann, J.~M. Henn, N.~A. Lo~Presti, V.~Mitev and P.~Wasser,
  \emph{{Analytic result for the nonplanar hexa-box integrals}},
  \href{http://dx.doi.org/10.1007/JHEP03(2019)042}{\emph{JHEP} {\bf 03} (2019)
  042}, [\href{https://arxiv.org/abs/1809.06240}{{\tt 1809.06240}}].

\bibitem{CaronHuot:2012ab}
S.~Caron-Huot and K.~J. Larsen, \emph{{Uniqueness of two-loop master
  contours}}, \href{http://dx.doi.org/10.1007/JHEP10(2012)026}{\emph{JHEP} {\bf
  10} (2012) 026}, [\href{https://arxiv.org/abs/1205.0801}{{\tt 1205.0801}}].

\bibitem{Remiddi:1999ew}
E.~Remiddi and J.~A.~M. Vermaseren, \emph{{Harmonic polylogarithms}},
  \href{http://dx.doi.org/10.1142/S0217751X00000367}{\emph{Int. J. Mod. Phys.}
  {\bf A15} (2000) 725--754}, [\href{https://arxiv.org/abs/hep-ph/9905237}{{\tt
  hep-ph/9905237}}].

\bibitem{Ahmed:2019qtg}
T.~Ahmed, J.~Henn and B.~Mistlberger, \emph{{Four-particle scattering
  amplitudes in QCD at NNLO to higher orders in the dimensional regulator}},
  \href{http://dx.doi.org/10.1007/JHEP12(2019)177}{\emph{JHEP} {\bf 12} (2019)
  177}, [\href{https://arxiv.org/abs/1910.06684}{{\tt 1910.06684}}].

\bibitem{Henn:2019rgj}
J.~M. Henn and B.~Mistlberger, \emph{{Four-graviton scattering to three loops
  in $ \mathcal{N}=8 $ supergravity}},
  \href{http://dx.doi.org/10.1007/JHEP05(2019)023}{\emph{JHEP} {\bf 05} (2019)
  023}, [\href{https://arxiv.org/abs/1902.07221}{{\tt 1902.07221}}].

\end{thebibliography}\endgroup
\bibliographystyle{JHEP}
\end{document}